\documentclass[journal]{IEEEtran}

\usepackage{enumitem}

\usepackage{graphicx}      
\usepackage{float}
\usepackage{caption}
\usepackage{subcaption}
\usepackage{epsfig} 
\usepackage{mathptmx} 
\usepackage{times} 
\usepackage{amsmath} 
\usepackage{amssymb}  
\usepackage{amsfonts}
\usepackage{xspace}
\usepackage{pifont}

\usepackage{comment}
\usepackage{todonotes} 
\presetkeys{todonotes}{inline,fancyline}{}
\usepackage{color}

\usepackage{cite}
\usepackage[multiple]{footmisc}
\usepackage{hyperref}

\usepackage{balance} 

\usepackage{algorithm}
\usepackage[noend]{algpseudocode}

\usepackage{tikz}
\usetikzlibrary{automata,arrows}
\usetikzlibrary{matrix}
\usetikzlibrary{patterns}

\newcommand*\circled[1]{\tikz[baseline=(char.base)]{
            \node[shape=circle,draw,inner sep=0.8pt] (char) {#1};}}
\newcommand{\false}{\ensuremath{\mathit{false}}\xspace}
\newcommand{\true}{\ensuremath{\mathit{true}}\xspace}

\newcommand{\SigmaFor}{\ensuremath{\Sigma_{\mathit{for}}}\xspace}
\newcommand{\RealTime}{\ensuremath{\mathbb{R}_{\geq 0}}\xspace}
\newcommand{\Sigmauc}{\ensuremath{\Sigma_{\mathit{uc}}}\xspace}
\newcommand{\pred}{\ensuremath{\mathit{pred}}\xspace}
\newcommand{\Preds}{\ensuremath{\mathit{Preds}}\xspace}
\newcommand{\Reg}{\ensuremath{\mathit{Reg}}\xspace}
\newcommand{\NBP}{\ensuremath{\mathit{NBP}}\xspace}
\newcommand{\BSP}{\ensuremath{\mathit{BSP}}\xspace}
\newcommand{\TSCS}{\ensuremath{\mathit{TSCS}}\xspace}

\newcommand{\app}{\ensuremath{\mathit{app}}\xspace}
\newcommand{\inE}{\ensuremath{\mathit{in}}\xspace}
\newcommand{\out}{\ensuremath{\mathit{out}}\xspace}
\newcommand{\exit}{\ensuremath{\mathit{exit}}\xspace}
\newcommand{\down}{\ensuremath{\mathit{down}}\xspace}
\newcommand{\lowerE}{\ensuremath{\mathit{lower}}\xspace}
\newcommand{\up}{\ensuremath{\mathit{up}}\xspace}
\newcommand{\raiseE}{\ensuremath{\mathit{raise}}\xspace}

\makeatletter

\newcommand{\tinytodo}[2][]
{\todo[caption={#2}, size=\scriptsize, #1]{\renewcommand{\baselinestretch}{0.9}\selectfont#2\par}}

\newcommand{\AR}[1]{\tinytodo[backgroundcolor=yellow]{#1}}
\newcommand{\MR}[1]{\tinytodo[backgroundcolor=green]{#1}}
\newcommand{\Q}[1]{\tinytodo[backgroundcolor=orange]{#1}}

\newcommand{\Rmnum}[1]{\expandafter\@slowromancap\romannumeral #1@}

\makeatother
\newtheorem{defn}{Definition}
\newtheorem{property}{Property}

\newtheorem{example}{Example}
\newtheorem{theorem}{Theorem} 
\newtheorem{lemma}{Lemma}
\newtheorem{corollary}{Corollary}
\newtheorem{proof}{Proof}
\newtheorem{remark}{Remark}

\begin{document}
\title{Supervisory Control Synthesis of Timed Automata Using Forcible Events}
%
%
%

\author{Aida~Rashidinejad~\IEEEmembership{} \and
        Michel~Reniers~\IEEEmembership{} \and
        Martin~Fabian~\IEEEmembership{}

\thanks{This research has received funding from the European Union’s Horizon 2020 Framework Programme for Research and Innovation under grant agreement no 674875.}

\thanks{Aida Rashidinejad and Michel Reniers are with the Control Systems Technology Group, Department of Mechanical Engineering, Eindhoven University of Technology, P.O.Box 513, 5600 MB\ Eindhoven, The Netherlands (e-mail: \{a.rashidinejad, m.a.reniers\}@tue.nl).}
\thanks{Martin Fabian is with Department of Electrical Engineering, Chalmers University of Technology, Sweden (e-mail: fabian@chalmers.se).}%
}

\maketitle


\begin{abstract}
Considering real-valued clocks in timed automata (TA) makes it a practical modeling framework for discrete-event systems. However, the infinite state space brings challenges to the control of TA.
To synthesize a supervisor for TA using the conventional supervisory control theory, existing methods abstract TA to finite automata (FA).
For many applications, the abstraction of real-time values results in an explosion in the state space of FA.
This paper presents a supervisory control synthesis algorithm directly applicable to the TA without any abstraction.
The plant is given as a TA with a set of uncontrollable events and a set of forcible events. Forcible events can preempt the passage of time when needed.
The synthesis algorithm works by iteratively strengthening the guards of edges labeled by controllable events and invariants of locations where the progression of time can be preempted by forcible events.
The synthesized supervisor, which is also a TA, is guaranteed to be controllable, maximally permissive, and results in a nonblocking and safe supervised plant.   
\end{abstract}


\IEEEpeerreviewmaketitle


\section{Introduction}
\label{section:intro}
\IEEEPARstart{S}{upervisory}  control theory (SCT) was first introduced by Ramadge-Wonham to control discrete-event systems (DES)~\cite{ramadge1989control}. SCT provides a synthesis method resulting in a supervisor that restricts the plant behavior towards a given set of desired behavior. Moreover, the synthesized supervisor satisfies the controllability, nonblockingness, and maximal permissivesness properties~\cite{wonham2015supervisory}.

DES, such as communication networks, manufacturing and traffic systems,
are typically modeled using finite automata (FA). To provide a compact representation of complex and large DES, FA have been further extended with discrete variables to extended finite automata (EFA)~\cite{EFAmodeling}. In EFA, transitions are labeled by events and associated with constraints on variables (guards), where variables may be updated after the occurrence of an event~\cite{EFAmodeling}.

The dynamics of DES depend entirely on the ordering of the event occurrences, and so are independent of time~\cite{cassandras2009introduction}.
However, the control of many applications needs to be able to include timing information in modeling DES.
Imagine a system that needs to be controlled over a distance, due to being located in a hazardous or unreachable environment.
To control such systems, the concept of networked supervisory control has been introduced~\cite{Lin:14,Rashidinejad18}.

Networked control of systems introduces communication delays that are unavoidable and have a high impact on the system performance~\cite{Heemels:10}.
To consider the effects of communication delays, the DES model must include timing information of event occurrences as well as the ordering of them. 
For this purpose, the concepts of timed discrete-event systems (TDES), and timed automata (TA) have been introduced in~\cite{Wonham:94} and~\cite{alur1994theory}, respectively.
TDES and TA are known as real-time discrete-event systems (RTDES), which are modeled not only based on the ordering of events, but also based on timing constraints on events~\cite{khoumsi2002supervisory}.

TDES incorporate discrete time in modeling DES.
A TDES is generally a DES in which the execution of each event, called \emph{active} event, is restricted within a lower and an upper time bound specified for the event.
It is assumed that a digital clock exists in the system, and so the TDES is modeled as a FA that includes a specific event, called \emph{tick}, indicating the passage of a unit of time.
The event \emph{tick} is generally an uncontrollable event as it spontaneously occurs in the system, and so it cannot be disabled by a supervisor. However, it is assumed that \emph{tick} is \emph{preemptable} by a subset of active events, called \emph{forcible} events.
Taking the nature of \emph{tick} into account, SCT of DES, has been modified for TDES in~\cite{Wonham:94}.
Moreover, like DES, the model of TDES has been extended with discrete variables into timed extended finite automata (TEFA)~\cite{miremadi2015symbolic}. 

TA incorporate dense-time in modeling DES~\cite{alur1994theory}. A TA consists of a finite set of locations and a finite set of real-valued clocks~\cite{dubey2009discussion}. To each location, a clock constraint is associated, called an \emph{invariant}, determining the time that the system is allowed to stay in that location. Each edge between two locations is labeled by an event, the clock constraint associated to that event called the \emph{guard}, and the set of clocks that are reset to zero, called the \emph{reset}, by the occurrence of that event.

Compared to TDES, a TA brings a more natural modeling framework for real-life applications because 1) it considers real-time, and so it copes with the state space explosion problem introduced by discrete time; this is especially important for systems with various time scales. And 2) it easily allows events to have multiple and different timing constraints, rather than specifying the time of each event occurrence by fixed lower and upper bounds.


The control of TA is challenging due to the clock variables, making the state space of TA infinite.
To overcome this problem,  existing approaches abstract TA into FA, and apply supervisory control synthesis on the abstracted result~\cite{wong1991control,tripakis1999fly,maler1995synthesis}.
In general, the synthesis approaches can be divided into the following categories: 1) game-based (reactive) synthesis, and 2) the synthesis method proposed by Ramadge-Wonham, which is referred to as RW-based synthesis in this paper.
Game-based (reactive) synthesis of TA has been investigated in~\cite{maler1995synthesis,asarin1998controller,tripakis1999fly,cassez2005efficient}, and it has also been implemented in tools such as  UPPAAL-TIGA~\cite{behrmann2007uppaal,maler1995synthesis}.
Game-based synthesis and RW-based synthesis mainly differ in satisfying maximal permissiveness. While RW-based synthesis provides a unique maximally permissive supervisor, game-based synthesis gives a winning strategy if it exists, which is not necessarily the maximally permissive solution~\cite{ehlers2017supervisory}.
In this paper, we focus on RW-based synthesis as we  
want to achieve a maximally permissive, controllable, and nonblocking supervisor.

RW-based supervisor synthesis
of TA was first investigated in~\cite{wong1991control}, where the plant is first abstracted into an FA (region graph) using region-based abstraction from~\cite{alur1994theory,wong1991control}.
Then, a supervisor is synthesized for the FA using  existing methods. Finally, to refine the abstraction, timing information is added to the FA supervisor.
For many applications, region-based abstraction results in a finite but a very large FA~\cite{khoumsi2002efficient,tripakis2001analysis}.

To overcome the state-space explosion problem of region-based abstraction, some state-space minimization methods have been proposed such as zone-based abstraction~\cite{alur1994theory}. These methods are mainly used for model checking and verification purposes as they do not provide sufficient information for supervisor synthesis~\cite{ouedraogo2010setexp}.

In~\cite{khoumsi2002efficient,ouedraogo2010setexp}, a transformation is introduced to obtain a minimal FA from a TA that is suitable for synthesis purposes.
The transformation is based on two special events; \emph{Set} and \emph{Exp}, where \emph{Set} represents the set and reset of a clock, and \emph{Exp} indicates the expiration of the clock. 
The \emph{SetExp}-transformation results in a minimal FA, for which a supervisor is synthesized using the concept of forcible events from TDES.
Preempting time using forcible events results in a more comprehensive solution as more events can be disabled if needed. 
However, it is currently unknown how to refine the synthesized supervisor (as an FA with \emph{Set} and \emph{Exp} events) to a TA (with these events translated into time constraints), and so the synthesis based on \emph{SetExp}-transformation is not satisfying.

Supervisory control of TA using forcible events is also investigated in~\cite{ICARCV20}, in which region-based abstraction is used to abstract a TA into an FA. For the FA, a synthesis algorithm is proposed. The synthesized supervisor is transformed back into a TA using a time-refinement technique. 
Although this method gives the supervisor as a TA, it still suffers from the state-space explosion problem caused by the abstraction.


This paper provides a supervisory control technique for TA such that:

\begin{itemize}
    \item no abstraction is needed to cope with the state-space explosion problem of some existing approaches,
    \item an algorithm is proposed that works with automata instead of languages to ease integration of an implementation in a tool set such as CIF or Supremica~\cite{CIF,Supremica},
    \item the RW-based synthesis is used so that the synthesized supervisor is maximally permissive, as well as controllable, and nonblocking,
    \item the concept of forcible events from TDES is used to provide a more comprehensive result, and
    \item to provide technical proofs, the notion of clock regions of timed automata is adapted in a specific way.
\end{itemize}


To the best of our knowledge, there is no work in the literature investigating TA RW-based synthesis without abstraction as we do here.
Our synthesis technique is  close to supervisory control synthesis for EFA. The main differences between EFA and TA are as follows: 1) an EFA deals with a set of variables belonging to a finite domain. However, a TA deals with clock variables, which belong to the infinite set of real-valued numbers, and
2) a TA includes location invariants that force the TA to leave the location before the invariant is violated. This is not the case in EFA.
Dealing with real-valued clock variables and location invariants make the synthesis of TA much more complex than the synthesis of EFA. Details are discussed throughout the paper. 

An earlier version of this work has been published in~\cite{Rashidinejad:20}.
Compared to~\cite{Rashidinejad:20}, this paper 1) provides the detailed proofs, 2) generalizes the approach for control requirements that are generally given as automata, and 3) applies the method to a
well-known case study.

The rest of the paper is organized as follows.
In Section~\ref{section:preliminaries}, the formal definition of TA and the relevant concepts are given.
Section~\ref{section:synthesis} presents the basic timed supervisory control (TSC) synthesis problem and the proposed solution. In Section~\ref{section:requirement}, the basic TSC synthesis problem is generalized to satisfy a given set of control requirements. 
To verify the results, the proposed method is applied to a rail road crossing system in Section~\ref{section:CaseStudy}.
Finally, Section~\ref{section:conclusion} concludes the paper.
To enhance readability, all technical lemmas and proofs are given in the appendices.

\section{Preliminaries}
\label{section:preliminaries}
A TA is an FA extended with a finite set of real-valued clocks. 
To model the timing behavior of TA, the accepting temporal conditions to switch between different modes (locations) or stay in the current one are represented by clock constraints~\cite{alur1994theory,Bengtsson2004}.

\begin{defn}[Clock Constraints~\cite{Bengtsson2004}]
\label{def:constraint}
Given a finite set of real-valued clocks $C$, $x\sim n$ and $x-y\sim n$ are atomic clock constraints for any $x,y\in C$, ${\sim}\in\{<,=,>\}$, and $n\in\mathbb{N}$. Clock constraints are defined as follows: any atomic clock constraint is a clock constraint, and for any two clock constraints $\varphi_1$ and $\varphi_2$, also $\varphi_1 \wedge \varphi_2$ and $\varphi_1 \vee \varphi_2$ are clock constraints.
\hfill$\blacksquare$
\end{defn}

Instead of writing $x-x=0$ with $x\in C$ as a clock constraint, we write $\true$. Similarly, $\false$ is written instead of $x-x>0$.

\begin{defn}[Clock Valuation]
\label{def:clock valuation}
Given a set of clocks $C$, a clock valuation $u: C\rightarrow \RealTime$ assigns a real value to each clock $x\in C$. 
\hfill$\blacksquare$
\end{defn}

Note that, initially, the valuation of each clock is $\textbf{0}$, where $\mathbf{0}$ denotes the clock valuation where all the clock variables have value 0.

A clock valuation $u$ satisfies a clock constraint $\varphi$, denoted $u\models \varphi$, whenever $\varphi$ is $\true$ for the values assigned by $u$ to each clock. 


\begin{defn}[Timed Automaton~\cite{alur1994theory}]
\label{def:TA}
A timed automaton is a 7-tuple $(C,L,\Sigma,E,L_m,L_0,I)$ where 
	
\begin{itemize}
	\item $C$ is a finite set of clocks with a non-negative real-value (from $\mathbb{R}_{\geq 0}$). The initial value of each  clock variable is always assumed to be 0,
	\item $L$ is a finite set of locations,
	\item $\Sigma$ is a finite set of events,
	\item $E$ is a finite set of edges with elements $e$ of the form $(l_s,\sigma,g,r,l_t)$ for which $l_s,l_t\in L$ are the source and target locations, respectively, $\sigma\in \Sigma$, $g$ is the guard  which is a clock constraint,
	and $r\subseteq C$ is the set of clocks to be reset to 0,
	\item $L_m\subseteq L$ is the set of marked locations,
	\item $L_0\subseteq L$ is the set of initial locations,
	\item $I$ is a function associating an invariant to each location $l\in L$. An invariant is a clock constraint that needs to be satisfied when the system is in the location. \hfill $\blacksquare$
\end{itemize}
\end{defn}
In~\cite{Bengtsson2004}, guards are generally given as clock constraints, but invariants are restricted to clock constraints that are downwards closed; $x<n$ or $x\leq n$.
In this work, similar to~\cite{alur1999timed}, both guards and invariants are allowed to be arbitrary clock constraints.


To clarify the problem and illustrate each step of the approach, the bus-pedestrian example from~\cite{brandin1994supervisory} is used throughout the paper.

\begin{example}[Bus-Pedestrian]
\label{ex:busped}
Imagine that a bus is headed directly for a pedestrian and will run over him at time $x = 2$ if he does not move. The pedestrian needs an amount of time $y = 1$ to realize his fate, after which he has the chance to jump out of the bus's path. If the pedestrian jumps before the bus passes, he is safe. Figure~\ref{fig:BP} gives the automata, representing the bus, the pedestrian, and the safe behavior of the system. The safe behavior is modeled in such a way that if the pedestrian jumps before the bus passes, then the system goes to a marked state. Otherwise, the system goes to a blocking state. 

\begin{figure}[h]
    \centering
    \begin{subfigure}[b]{0.4\columnwidth}
        \centering
        \begin{tikzpicture}[>=stealth', shorten >=1pt, auto, node distance=3.5cm, scale=.75, transform shape, align=center,state/.style={circle, draw, minimum size=1.2cm,font=\small}]

    \node[initial,initial text={},state]           (A)                                    {$a$ \\$x\leq 2$};
    \node[state,accepting]         (B) [right of=A]                       {$g$};

    \path[->] (A) edge [above,dashed]   node [align=center]  {\small{$x=2$}\\ \small{$pass$}} (B);
    \end{tikzpicture}
    \caption{Bus}
    \end{subfigure}
\hspace{1cm}
    \begin{subfigure}[b]{0.45\columnwidth}
    \centering
    \begin{tikzpicture}[>=stealth', shorten >=1pt, auto, node distance=3.5cm, scale=.75, transform shape, align=center,state/.style={circle, draw, minimum size=1.2cm,font=\small}]

    \node[initial,initial text={},state]           (A)                                    {$r$};
    \node[state,accepting]         (B) [right of=A]                       {$c$};

    \path[->] (A) edge [above]   node [align=center]  {\small{$y\geq 1$}\\ \small{\underline{$jump$}}} (B);
    \end{tikzpicture}
    \caption{Pedestrian}
    \end{subfigure}
\hspace{1cm}
   \begin{subfigure}[b]{1\columnwidth}
    \centering
    \begin{tikzpicture}[>=stealth', shorten >=1pt, auto, node distance=3.5cm, scale=.75, transform shape, align=center,state/.style={circle, draw, minimum size=1.2cm,font=\small}]

    \node[initial,initial text={},state]           (A)                                    {$0$};
    \node[state]         (B) [right of=A]                       {$1$};
    \node[state,accepting]         (C) [right of=B]                       {$2$};
    \node[state]         (D) [below of=B]                       {$\bot$};

    \path[->] (A) edge [above]   node [align=center]  {\small{\underline{$jump$}}} (B)
    (B) edge [above,dashed]   node [align=center]  {\small{$pass$}} (C)
    (A) edge [left,dashed]   node [align=center]  {\small{$pass$}} (D)
    (C) edge [right,dashed]   node [align=center]  {\small{$pass$}} (D)
    (D) edge [loop left, dashed]   node [align=center]  {\small{$pass$}} (D);
    \end{tikzpicture}
    \caption{Safe behavior}
    \end{subfigure}
\caption{Plant automata from Example~\ref{ex:busped}.}
\label{fig:BP}
\end{figure}
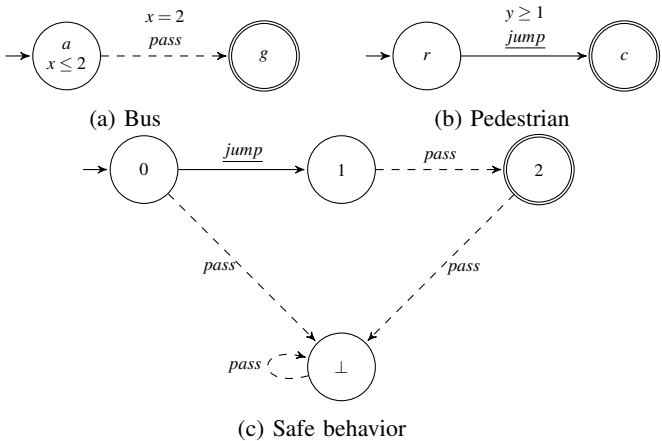
\end{example}

For TA, we frequently use the following notations:
\begin{itemize}
    \item the notation $.$ is used to refer to an element of a tuple. For instance, $e.\sigma$ refers to $\sigma$ from the edge $e\in E$.
    \item the notation $\pred^{\uparrow\delta}$, for a predicate $\pred$ and the increase $\delta\in\RealTime$,
    replaces all occurrences of the variables $x\in C$ by $x+\delta$. For instance, $(x\geq 3)^{\uparrow\delta}$ gives $x+\delta\geq 3$.   
    \item the notation $\pred[r]$, for a predicate $\pred$ and a reset $r$. 
    The meaning of this notation is a predicate in which all occurrences of clock variables from $r$ are replaced by zero. 
    \item the notation $\Preds(C)$, to indicate the set of all predicates over the clock variables.
    \item The notation $P$ stands for the natural projection operator as defined in~\cite{cassandras2009introduction}; given a language $L\subseteq\Sigma^*$ and an event set $\Sigma'\subseteq \Sigma$: $P_{\Sigma'}(L):=\{w'\in\Sigma'^*\mid \exists w\in L, P_{\Sigma'}(w)=w'\}$.
\end{itemize}

In this paper, we only deal with \emph{deterministic} TA.

\begin{defn}[Deterministic TA~\cite{alur1994theory}]
\label{dfn:deterministic}
A timed automaton $(C,L,\Sigma,E,L_m,L_0,I)$ is deterministic if it has only one initial location $L_0=\{l_0\}$,
and for any pair of edges $e_1,e_2\in E$, with the same source location ($e_1.l_s=e_2.l_s$) and labeled by the same event ($e_1.\sigma=e_2.\sigma$),
the clock constraints are mutually exclusive ($e_1.g\land e_2.g=\mathit{false}$).
\hfill $\blacksquare$
\end{defn}

From now on, we only use TA with a single initial location $l_0$ and consequently represent them by $(C,L,\Sigma,E,L_m,l_0,I)$.

In the examples, TA are depicted graphically. The locations are represented by circles and the edges by arrows from the source location to the target location, labelled with the event, the guard and the reset. The reset of a clock $x \in r$ is denoted by $x:=0$.
Invariants of locations are indicated inside the locations. Absence of an invariant in a location represents the invariant that always holds. The initial location is depicted by a dangling incoming arrow, and the marked locations by double circles. 

\begin{defn}[Sub-automaton of a TA]
\label{def:subA}
Given a TA $A=(C,L,\Sigma,E,L_m,l_0,I)$, a TA $B=(C,L',\Sigma,E',L'_m,l'_0,I')$ is a sub-automaton of $A$, denoted $B\subseteq A$, if 

\begin{itemize}
    \item $L'\subseteq L$,
    \item for all $(l_s,\sigma,g',r,l_t)\in E': (l_s,\sigma,g,r,l_t) \in E$ for some $g$ such that $g' \Rightarrow g$,
    \item $L'_m = L_m \cap L'$,
    \item $l'_0=l_0$, and
    \item for all $l\in L'$: $I'(l) \Rightarrow I(l)$.
\hfill$\blacksquare$
\end{itemize}
\end{defn}

Applications are typically modeled by a network of automata, where each automaton represents a single component or subsystem; compare Figure~\ref{fig:BP}. A single automaton representing the network of automata can then be generated as the synchronous product of the constituent automata.

In~\cite{alur1994theory,Bengtsson2004}, synchronous product of TA is defined under the assumption that the two TA do not share any clock variable. This assumption is relaxed here, and the synchronous product is generalized for TA with shared set of clocks. To do so, we are inspired from the synchronous product of two EFA as defined in~\cite{EFAmodeling}.



\begin{defn}[Synchronous Product of TA]
\label{dfn:SynchProduct}
The synchro\-nous product of two TA $G_1 = (C_1,L_1,\Sigma_1,E_1,L_{1m},l_{10},I_1)$ and $G_2 = (C_2,L_2,\Sigma_2,E_2,L_{2m},l_{20},I_2)$, 
is given by $G_1||G_2= (C_1 \cup C_2, L_1 \times L_2,\Sigma_1 \cup \Sigma_2,E_p,L_{1m} \times L_{2m} ,(l_{10},l_{20}), I_p)$, where for each $l_1\in L_1$ and $l_2\in L_2$, $I_p(l_1,l_2) = I_1(l_1) \wedge I_2(l_2)$ and each edge in $E_p$ is as follows: 
		\begin{itemize}
			\item $\sigma \in \Sigma_1 \setminus \Sigma_2$, then for every $(l_{s1}, \sigma, g_1, r_1, l_{t1})\in E_1$ and $l_{2} \in L_2$, $((l_{s1},l_{2}), \sigma, g_1, r_1,  (l_{t1},l_{2}))\in E_p$
        	\item $\sigma \in \Sigma_2 \setminus \Sigma_1$, then for every $(l_{s2}, \sigma, g_2, r_2, l_{t2})\in E_2$ and $l_{1} \in L_1$, $((l_{1},l_{s2}), \sigma, g_2, r_2,  (l_{1},l_{t2}))\in E_p$.
			\item $\sigma \in \Sigma_1 \cap \Sigma_2$, then for every $(l_{s1}, \sigma, g_1, r_1, l_{t1})\in E_1$ and $(l_{s2}, \sigma, g_2, r_2, l_{t2})\in E_2$, $((l_{s1},l_{s2}), \sigma, g_1\wedge g_2, r_1 \cup r_2, (l_{t1},l_{t2}))\in E_p$. 
			\hfill$\blacksquare$
		\end{itemize}
\end{defn}

\begin{figure}[h]
\centering
    \begin{tikzpicture}[>=stealth', shorten >=1pt, auto, node distance=3.5cm, scale=.75, transform shape, align=center,state/.style={circle, draw, minimum size=1.2cm,font=\small}]

    \node[initial,initial text={},state]           (A)                                    {$(a,r,0)$ \\$x\leq 2$};
    \node[state]         (B) [right of=A]                       {$(g,r,\bot)$};
    \node[state]         (C) [below of=A]                     {$(a,c,1)$ \\$x\leq 2$};
    \node[state,accepting]         (D) [right of=C]                       {$(g,c,2)$};

    \path[->] (A) edge [above,dashed]   node [align=center]  {\small{$x=2$}\\ \small{$pass$}} (B)
    (A) edge [right]   node [align=center]  {\small{$y\geq 1$}\\ \small{\underline{$jump$}}} (C)
    (C) edge [above,dashed]   node [align=center]  {\small{$x=2$}\\ \small{$pass$}} (D);
    \end{tikzpicture}
\caption{Synchronous product of the TA from Example~\ref{ex:busped}.}
\label{fig:BP-synch}
\end{figure}
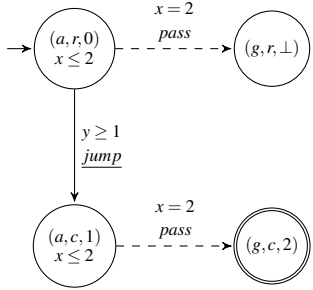

For the bus-pdestrian example, the synchronous product of the bus, pedestrian and the safe behavior automata is shown in Figure~\ref{fig:BP-synch}.

Every TA has an underlying semantic graph~\cite{alur1994theory,tripakis2001analysis}. 


\begin{defn}[Semantic Graph]
	\label{def:SG}
The semantic graph of a TA $G=(C,L,\Sigma,E,L_m,l_0,I)$, is 
a labeled graph with a set of states $X\subseteq L \times (C \rightarrow\RealTime)$, consisting of a location $l$ and a clock valuation $u$ such that $(l,u)\in X$ iff $u\models I(l)$. The initial state is $(l_0,\mathbf{0})$ if $\mathbf{0}\models I(l_0)$. Otherwise, the semantic graph is undefined.  The semantic graph has the following transitions:
	\begin{itemize}
		\item event transition: from state $(l_s,u_s)$ to state $(l_t,u_s[r])$ labeled by event $\sigma$ if there is an edge $e=(l_s,\sigma,g,r,l_t)$ such that $u_s\models g$, and $u_s[r] \models I(l_t)$.
		\item time transition: from state $(l,u)$ to state  $(l, u+\Delta )$
		labeled with delay $\Delta\in\RealTime$ if $u+\delta\models I(l)$ for any $\delta$ such that $0 \leq \delta \leq \Delta$. Note that for a valuation $u$ and a real value $\delta$, $u+\delta$ denotes the clock valuation with $(u+\delta)(x) = u(x) + \delta$ for each clock $x\in C$.
	\end{itemize}

Moreover, states $(l,u)$ in the semantic graph with $l\in L_m$ (regardless of the clock valuation $u$) are marked.
A word $w$ in the semantic graph of $G$ is a finite sequence of labels; $w\in(\Sigma\cup\mathbb{R}_{\geq 0})^*$ with $\varepsilon$ denoting the empty sequence.
A state in the semantic graph of $G$ is called reachable if it can be reached from the initial state via a word. The language of $G$, indicated by $L(G)$, is the set of all words in its semantic graph starting from the initial state.
Note that for any $G'\subseteq G$: $L(G')\subseteq L(G)$.
\hspace*{\fill} $\blacksquare$
\end{defn}

Note that since a TA is allowed to have arbitrary clock constraints as invariants, it may be the case that $\mathbf{0}\not\models I(l_0)$. This may happen regarding modeling issues, or through synthesis, where in the latter case, synthesis actually does not result in a supervisor.

Based on the semantic graph, some relevant notions for timed automata are defined.

\begin{defn}[Nonblockingness]
		\label{def:Nonblockingness}
A state in a semantic graph is nonblocking if there exists a path 
leading from that state to a marked state, i.e., a state $(l_t, u_t)$ with $l_t\in L_m$. 
A TA is nonblocking if all of the reachable states in its semantic graph are nonblocking.
	\hfill $\blacksquare$
	\end{defn}

In the rest of the paper, the plant is given as a TA $G$ represented by $(C,L,\Sigma_G,E_G,L_m,l_0,I_G)$.
It is assumed that all events are observable. However, not all of the events might be controllable.
The set of events $\Sigma_G$ is assumed to be partitioned into a set of uncontrollable events $\Sigmauc$ and a set of controllable events $\Sigma_c=\Sigma_G\setminus\Sigmauc$.
Uncontrollable events are events that occur spontaneously in the plant such as disturbances or sensor readings. Controllable events are signals sent to the actuators. In figures of TA, edges labelled by uncontrollable events are indicated by dashed lines, and edges labelled by controllable events are indicated by solid lines.
Time passage is uncontrollable by nature. However, it may be preempted by execution of a forcible event 
$\sigma_f \in \SigmaFor$, where $\SigmaFor\subseteq\Sigma_G$ (forcible events are underlined in figures). Consequently, considering the semantic graph of a TA, a time transition enabled at a state is considered uncontrollable by default, unless there is also a forcible event transition enabled at that state. Then, the time transition is said to be \emph{preemptable}. Note that a forcible event can be controllable or uncontrollable as discussed in~\cite{wonham2015supervisory}.
For the bus-pedestrian example, the event $\mathit{pass}$ is uncontrollable, and the event $\mathit{jump}$ is controllable and forcible.

The following definition of \emph{controllability} for TA with forcible events, is inspired from~\cite{brandin1994supervisory}. 

\begin{defn}[Controllability of TA with Forcible Events]
\label{dfn:controllability}
Given a plant $G$ with uncontrollable events $\Sigmauc$, and forcible events $\SigmaFor$, a TA $S$ is \emph{controllable} w.r.t.\ $G$ if for all $w\in L(S||G)$ and $\sigma\in\Sigmauc\cup\mathbb{R}_{\geq 0}$, whenever $w\sigma\in L(G)$:
\begin{enumerate}
    \item\label{defCtrlStandard} $w\sigma\in L(S||G)$, or
    \item\label{defCtrlForcible} $\sigma\in\mathbb{R}_{\geq 0}$ and $w\sigma' \in L(S||G)$ for some $\sigma'\in\SigmaFor$. 
    
Property~\eqref{defCtrlStandard} above is the standard controllability property; $S$ cannot disable uncontrollable events that $G$ may generate. However, if a forcible event is enabled, this may preempt the time event, which is captured by Property~\eqref{defCtrlForcible}.
\hfill $\blacksquare$
\end{enumerate}
\end{defn}


A supervisor $S$ is called \emph{proper} for a plant $G$ whenever $S$ is controllable w.r.t.\ $G$, and the supervised plant $S||G$ is nonblocking.

\begin{defn}[Maximal Permissivenesss]
\label{dfn:MaxPer}
A proper supervisor $S$ is \emph{maximally permissive} for a plant $G$, whenever $S$ preserves the largest admissible behavior of $G$ compared to any other proper supervisor $S'$; for any proper $S'$: $L(S'||G)\subseteq L(S||G)$.
\hfill$\blacksquare$
\end{defn}


As stated in~\cite{alur1994theory}, the clock valuations of a TA $G$ can be divided into a finite set of clock regions using the definition of region equivalence.
Here, we introduce extended clock regions of a TA $G$, denoted $R_G$.

\begin{defn}[Extended Clock Regions of TA]
\label{def:clock region}
Consider a TA $G$ with a set of clocks $C$ where the the clock ceiling function, $k:C\rightarrow \mathbb{N}$ gives the largest natural number that a clock $x\in C$ is bounded to by guards or invariants. 
Each clock region $r_G\in R_G$ is specified by:

\begin{enumerate}
    \item \label{item1} for each clock $x\in C$, a single clock constraint of one of the following forms:
\begin{itemize}    
    \item $x=n$ for some $n \in \{ 0,\ldots , k(x)\}$, 
    \item $n-1<x<n$ for some $n \in \{ 1,2,\ldots,k(x)\}$, or 
    \item $x>k(x)$
    \end{itemize}
    
    \item for any two different clocks $x,y\in C$, a single clock constraint of one of the following forms:
    \begin{itemize}
        \item $y-x+k(x) = q$ for some $q \in \{ 0,\ldots, k(x)+k(y)\}$,
        \item $q-1 < y-x+k(x) < q$ for some $q \in \{ 1,\ldots, k(x)+k(y)\}$,
        \item $y-x+k(x) < 0$, or
        \item $y-x+k(x) > k(x) + k(y)$ \hfill$\blacksquare$
    \end{itemize}
\end{enumerate}
\end{defn}

Note that $k(x)$ does not restrict the value of the clock variable $x$; it only gives the largest number that $x$ is bounded to by guards or invariants. Considering Figure~\ref{fig:BP-synch}, $k(x)=2$. However, in location $(g,r,\bot)$, the value of $x$ can grow to any real number larger than or equal to 2.




\begin{example}
\label{exp:ECR}
Figure~\ref{fig:Alur-Example} depicts the extended clock regions for a TA with two clock variables $x,y$, where $k(x)=2$ and $k(y)=1$. The clock regions given for the same example in~\cite{alur1994theory} are indicated in black. 


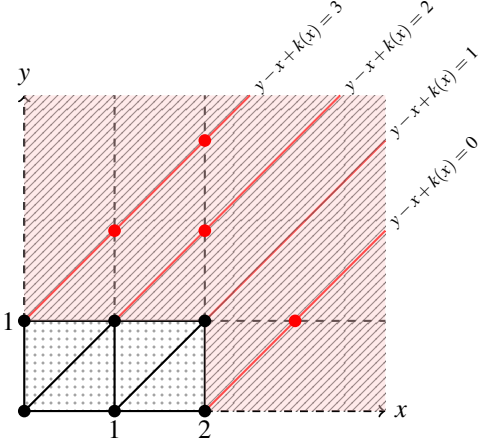
\begin{figure}[h]
\centering
\begin{tikzpicture}[scale=1.2]
    \draw [thick] (0,0) -- (0,1);
    \draw [thick, ->,dashed] (0,1) -- (0,3.5)
        node [above, black] {$y$};
    
    \draw [thick] (0,0) -- (2,0);
    \draw [thick, ->,dashed] (2,0) -- (4,0)
        node [right, black] {$x$}; 

    \node [left] at (0,1) {$1$}; 
    \node [below] at (1,0) {$1$}; 
    \node [below] at (2,0) {$2$}; 

    \draw [draw=black,thick] (0,1) -- (2,1);
    \draw [draw=black,thick,dashed] (2,1) -- (4,1);
    \draw [draw=black,thick] (1,0) -- (1,1);
    \draw [draw=black,thick,dashed] (1,1) -- (1,3.5);
    \draw [draw=black,thick] (2,0) -- (2,1);      
    \draw [draw=black,thick,dashed] (2,1) -- (2,3.5);

    \draw [draw=red,thick,font=\scriptsize] (0,1) -- (2.5,3.5) node[right,rotate=45] {$y-x+k(x)=3$}; 
    \draw [draw=black,thick,font=\scriptsize]  (0,0) -- (1,1) node[right,rotate=45] {}; 
    \draw [draw=red,thick,font=\scriptsize]  (1,1) -- (3.5,3.5) node[right,rotate=45] {$y-x+k(x)=2$};
    \draw [draw=black,thick,font=\scriptsize] (1,0) -- (2,1) node[right,rotate=45] {};
    \draw [draw=red,thick,font=\scriptsize] (2,1) -- (4,3) node[right,rotate=45] {$y-x+k(x)=1$};    
    \draw [draw=red,thick,font=\scriptsize] (2,0) -- (4,2) node[right,rotate=45] {$y-x+k(x)=0$};



    
    \draw[fill=gray,opacity=0.6,pattern=dots]  (0,0) -- (1,1) -- (1,0) -- cycle;    \draw[fill=gray,opacity=0.6,pattern=dots]  (0,0) -- (1,1) -- (0,1) -- cycle;
    \draw[fill=gray,opacity=0.6,pattern=dots]  (1,0) -- (2,1) -- (1,1) -- cycle;
    \draw[fill=gray,opacity=0.6,pattern=dots]  (1,0) -- (2,1) -- (2,0) -- cycle;

    \draw[fill=gray,opacity=0.8,pattern=north east lines,draw=none]  (2,1) -- (2,3.5) -- (4,3.5) -- (4,1);
    \draw[fill=gray,opacity=0.8,pattern=north east lines,draw=none]  (0,1) -- (0,3.5) -- (2,3.5) -- (2,1);
    \draw[fill=gray,opacity=0.8,pattern=north east lines,draw=none]  (2,0) -- (2,1) -- (4,1) -- (4,0);  

    \draw[fill=red!20!,opacity=0.4,draw=none]  (0,1) -- (2.5,3.5) -- (4,3.5) -- (4,2) -- (3,1);
    \draw[fill=red!20!,opacity=0.4,draw=none]  (2,0) -- (2,1) -- (3,1);
    
    \draw[fill=red!20!,opacity=0.4,draw=none]  (0,1) -- (0,3.5) -- (2.5,3.5) ;
    \draw[fill=red!20!,opacity=0.4,draw=none]  (2,0) -- (4,2) -- (4,0);
    
    \fill[black!100!] (0,0) circle (0.07cm);
    \fill[black!100!] (0,1) circle (0.07cm);
    \fill[black!100!] (1,0) circle (0.07cm);
    \fill[black!100!] (1,1) circle (0.07cm);
    \fill[black!100!] (2,0) circle (0.07cm);
    \fill[black!100!] (2,1) circle (0.07cm);

    \fill[red!100!] (1,2) circle (0.07cm);
    \fill[red!100!] (2,2) circle (0.07cm);
    \fill[red!100!] (3,1) circle (0.07cm);
    \fill[red!100!] (2,3) circle (0.07cm);
    
\end{tikzpicture}
    \caption{Extended clock regions from Example~\ref{exp:ECR}.}
    \label{fig:Alur-Example}
\end{figure}
\end{example}

We call a clock region \emph{unbounded} (dashed areas/lines in Figure~\ref{fig:Alur-Example}) if it is related to $x>k(x)$ for some $x\in C$. Otherwise, the region is called \emph{bounded} (dotted areas/solid lines in Figure~\ref{fig:Alur-Example}).
Note that although the number of the extended clock regions is more than the number of clock regions, it is still finite because the set of clock regions is finite (see~\cite{alur1994theory} for details), and the extended clock regions include all the bounded regions from the set of clock regions, and it partitions each unbounded region into a finite number of new regions.
For instance, in Example~\ref{exp:ECR}, $0<x<1, y>1$ is an unbounded region that is partitioned into new regions as $0<x<1, y>1, y-x+k(x)<3$; $0<x<1, y>1, y-x+k(x)=3$; and $0<x<1, y>1, y-x+k(x)>3$.

\begin{defn}[$G$-Clock Constraint]
\label{defn:$G$-cc}
Consider a plant $G$ with a set of clocks $C$, the clock ceiling function $k:C\rightarrow \mathbb{N}$, and the set of regions $R_G$.
A clock constraint $\varphi$ is called a \emph{$G$-clock constraint} whenever all the atomic constraints of $\varphi$ are bounded by $k(x)$ for all $x\in C$.
\hfill $\blacksquare$
\end{defn}

Clearly, for any two $G$-clock constraints $\varphi_1$ and $\varphi_2$, $\varphi_1\wedge\varphi_2$ and $\varphi_1\vee\varphi_2$ are $G$-clock constraints.

Based on the extended clock regions, we are now able to discriminate the regions that satisfy a $G$-clock constraint. Let us consider Example~\ref{exp:ECR} again. Given a $G$-clock constraint $\varphi=x-y>2$, there does not exist a set of clock regions satisfying $\varphi$ based on the definition of clock regions in~\cite{alur1994theory}. However, considering Definition~\ref{def:clock region}, $y=0,x>2,y-x+k(x)<0$; $0<y<1,x>2,y-x+k(x)<0$; $y=1,x>2,y-x+k(x)<0$; and $y>1,x>2,y-x+k(x)<0$ are the extended clock regions satisfying $\varphi$. This discrimination will be the basis to prove the termination and correctness of the proposed algorithms.

Moreover, it is assumed that there exists a function $Z$ mapping a $G$-clock constraint $\varphi$ to the maximal set of regions from $R_G$ such that for any region $r_G \in Z(\varphi)$, and for any valuation $u$ represented by $r_G$, denoted $u\in r_G$, $u\models \varphi$.
For any two $G$-clock constraints $\varphi_1$ and $\varphi_2$, $Z$ necessarily satisfies the following properties:



\begin{itemize}
    \item $Z(\varphi_1\wedge \varphi_2)=Z(\varphi_1)\cap Z(\varphi_2)$ and $Z(\varphi_1\vee \varphi_2)=Z(\varphi_1)\cup Z(\varphi_2)$.
    \item Whenever $Z(\varphi_1)=Z(\varphi_2)$, $\varphi_1$ and $\varphi_2$ represent the same $G$-clock constraint.
\end{itemize}
Also, for the clock constraints represented by $\true$ and $\false$, the mapping gives $R_G$, and $\varnothing$, respectively.

\section{Basic TSC Synthesis}
\label{section:synthesis}

\subsection{Problem Formulation}
The \emph{Basic TSC Synthesis Problem} is defined as follows. Given a plant model $G$ as a TA, the objective is to synthesize a timed supervisor $S$, also as a TA, such that
\begin{itemize}
    \item $S$ is controllable w.r.t.\ $G$,
    \item $S||G$ is nonblocking,  and
    \item $S$ is maximally permissive w.r.t.\ $G$.
\end{itemize}

Considering the bus-pedestrian example, a supervisor is required to avoid reaching the blocking location $(g,r,\bot)$ in Figure~\ref{fig:BP-synch}.
The objective is to provide a supervisory control synthesis approach that does not need an abstraction. The synthesized supervisor should respect controllability  (Definition~\ref{dfn:controllability}), nonblockingness (Definition~\ref{def:Nonblockingness}), and be maximally permissive (Definition~\ref{dfn:MaxPer}).



To synthesize such a supervisor, it is needed to determine the states $(l,u)$ in the semantic graph that should be made unreachable, referred to as \emph{bad states}. These are the following types of states: 1) states that are blocking and should be avoided to take care of nonblockingness, and 2) states that lead to a bad state through an uncontrollable event or a time transition that cannot be preempted; these states should be avoided to respect controllability as well as nonblockingness. 
As the synthesis algorithm should not involve any abstraction, we need to determine the clock valuations for which a location of a TA is a bad state (in the semantic graph). 
For this purpose, we start by determining the clock valuations for which a location is nonblocking, referred to as the ``nonblocking predicate" of a location. 
Based on the nonblocking predicate, a ``bad state predicate" is associated to each location determining the clock valuations for which the location is mapped to a bad state in the semantic graph.

\subsection{Nonblocking Condition}
\label{subsection:NB}
Given a plant $G$, Algorithm~\ref{algo:NB} associates a nonblocking predicate $N(l)$ to each location $l\in L$. Initially (line~\ref{line:NB-initial}), $N^i(l)$ with $i=0$ is set to $I_G(l)$ if $l$ is a marked location, and to $\mathit{false}$ otherwise.
The nonblocking predicate of each location is updated (line~\ref{line:NB-update}) to $N^{i+1}(l)$ based on:

$\circled{1}$ the current nonblocking predicate $N^i(l)$,

$\circled{2}$ the condition for any outgoing edge $(l,\sigma,g,r,l')$ to lead to a nonblocking location (an event transition leading to a nonblocking state in the semantic graph), and

$\circled{3}$ the condition to stay (for some time delay $\delta\leq \Delta$) in a nonblocking location as long as the invariant is satisfied (represented by a time transition leading to a nonblocking state in the semantic graph).

This iterates until a fix-point is reached where the nonblocking predicate stays the same for all locations (line~\ref{line:NB-end}).

\renewcommand{\algorithmicrequire}{\textbf{Input: }}
\renewcommand{\algorithmicensure}{\textbf{Output: }}
\begin{algorithm}[h]
\caption{Nonblocking Predicate (\NBP)}
{\algorithmicrequire} {$G=(C,L,\Sigma_G,E_G,L_m,l_0,I_G)$}

{\algorithmicensure} {$N: L\rightarrow \Preds(C)$}
\label{algo:NB}
\begin{algorithmic}[1]
\State{$i:=0$}
\For{$l\in L$}
{$
	\label{eq:NInit}
	N^0(l) := 
	\begin{cases}
\textcolor{red}{I_G(l)}, & \text{if}\ l\in L_m,\\
	\mathit{false}, & \text{otherwise}
	\end{cases}
	$}
\EndFor
	\label{line:NB-initial}
\Repeat
{\For{$l\in L$}
{\begin{align*}
	\label{eq:NAlg} 
	&N^{i+1}(l) := \overbrace{N^i(l)}^{\circled{1}}\, \vee  \overbrace{\bigvee_{l \xrightarrow{\sigma, g, r} l'} (g\wedge \textcolor{red}{I_G(l')[r]} \wedge N^i(l')[r] )\, }^{\circled{2}} \vee \\
	& \overbrace{\textcolor{red}{\exists\Delta\; N^i(l)^{\uparrow\Delta} \wedge \forall\delta\leq \Delta\;
	I_G(l)^{\uparrow\delta}}}^{\circled{3}}
	\end{align*}}\label{line:NBupdate}
	\EndFor\label{line:NB-update}}
\State{$i:=i+1$}
\Until {$\forall l\in L~ N^{i}(l)=N^{i-1}(l)$}\label{line:NBterminate}
\label{line:NB-end}
\For {$l\in L$}
{$N(l):=N^i(l)$}
\EndFor
\end{algorithmic}
\end{algorithm}







Algorithm~\ref{algo:NB} follows the same steps as presented for the nonblocking predicate of EFA in~\cite{ouedraogo2011nonblocking} with the following adjustments (indicated in red in Algorithm~\ref{algo:NB}):
\begin{enumerate}
    \item The initial nonblocking condition for marked locations is set to the location invariant $I_G(l)$ instead of $true$. This is to take into account the invariants of the marked locations.
    \item In the update (line~\ref{line:NB-update}), the invariant of the target location is added to the second term to guarantee that the invariant of the target location is satisfied upon entering that location. 
    \item The third term is added to take into account the time transitions in the semantic graph of the TA that may be used for reaching a nonblocking state. 
\end{enumerate}

\begin{property}[\NBP Termination]
\label{prop:NBPtermination}
Given a plant $G$ with a set of locations $L$ and a set of regions $R_G$; Algorithm~\ref{algo:NB} terminates.
\end{property}
\begin{proof}
See Appendix~\ref{proof:NBPtermination}.
\hfill $\blacksquare$
\end{proof}


\begin{property}[$\NBP$ and Nonblocking States]
\label{prop:NBP}
Given a plant $G$ and $\NBP(G)$: for any $(l,u)$ in (the semantic graph of) $G$, $(l,u)$ is a nonblocking state  iff $u\models N(l)$, where $N=\NBP(G)$.
\end{property}

\begin{proof}
See Appendix~\ref{proof:NBP}.
\hfill $\blacksquare$
\end{proof}

\begin{example}[Nonblocking Predicate for Bus-Pedestrian]
\label{ex:BP-NB} 
Consider the bus-pedestrian from Example~\ref{ex:busped}. The result of Algorithm~\ref{algo:NB} is given in Table~\ref{table:BS-NB}.
The conditions for locations $(g,r,\bot)$ and $(g,c,2)$ are left out, as they are $\mathit{false}$ and $\mathit{true}$ respectively, for all iterations.
The condition $x \leq 2 \wedge (y \geq 1 \lor x-y \leq 1)$ is equivalent to $x \leq 2 \wedge x-y \leq 1$.



	\begin{table}[h]
		\centering
		\captionsetup{width=\columnwidth}
	    \caption{Nonblocking predicate for bus-pedestrian.}
	    \label{table:BS-NB}
		\begin{tabular}{|c||c|c|}
			\hline
			& \multicolumn{2}{c|}{\textbf{$N$}} \\ \hline
			\textbf{$i$} & \textbf{Loc $(a,r,0)$} & \textbf{Loc $(a,c,1)$} \\ \hline
			0 & $\false$ & $\false$ \\ \hline
			1 & $\false$ & $x = 2$ \\ \hline
			2 & $x = 2 \wedge y\geq 1$ & $x \leq 2$ \\ \hline
			3 & $x \leq 2\wedge (y\geq 1 \vee x-y\leq 1)$ & $x \leq 2$ \\ \hline\hline
			4 & $x \leq 2 \wedge x-y \leq 1$ & $ x \leq 2$ \\ \hline
		\end{tabular}
	\end{table}
\end{example}

\subsection{Bad State Condition}
\label{subsection:BS}

Given a plant $G$, and the nonblocking predicate computed by Algorithm~\ref{algo:NB}, Algorithm~\ref{algo:BS} associates a bad state predicate $B(l)$ to each location $l\in L$.

Initially, $B^i(l)$ with $i=0$ is set to the logical negation of $N(l)$ for each location $l\in L$ (line~\ref{line:BS-initial}) because these characterize the blocking states.
Then, the bad state predicate of each location  is updated to $B^{j+1}(l)$ (line~\ref{line:BS-update}) based on

$\circled{4}$ the previous bad state predicate  $B^j(l)$,

$\circled{5}$ the condition of any outgoing edge $(l,\sigma,g,r,l')$ labeled by an uncontrollable event $\sigma\in\Sigmauc$ to lead to a bad state (an uncontrollable event transition leading to a bad state in the semantic graph), and

$\circled{6}$ the condition of staying in a bad state for some time delay $\delta\leq \Delta$ as long as the invariant is satisfied for all the clock variables and while there is no forcible event able to preempt time for any $\delta'\leq \delta$ (an uncontrollable time transition leading to a bad state in the semantic graph).

This iterates until a fix-point is reached where the bad state predicate stays the same for all locations (line~\ref{line:BS-end}).

\renewcommand{\algorithmicrequire}{\textbf{Input: }}
\renewcommand{\algorithmicensure}{\textbf{Output: }}
\begin{algorithm}[h]
\caption{Bad State Predicate (\BSP)}
{\algorithmicrequire} {$G=(C,L,\Sigma_G,E_G,L_m,l_0,I_G), \NBP(G)$}

{\algorithmicensure} {$B: L\rightarrow \Preds(C)$}
\label{algo:BS}
\begin{algorithmic}[1]
\State{$j:=0$}
\For{$l\in L$}
{$B^0(l) := \neg N(l)$}
		\EndFor
	\label{line:BS-initial}
\Repeat
	\For{$l\in L$}
{\begin{align*}
&	B^{j+1}(l) :=  \overbrace{B^{j}(l)}^{\circled{4}} \vee \overbrace{\bigvee_{\substack{l \xrightarrow{\sigma, g, r} l'\\ \sigma\in \Sigmauc}} \big(g  \wedge \textcolor{red}{I_G(l')[r]}\wedge B^{j}(l')[r]\big)
\,}^{\circled{5}} \vee
\\
& \overbrace{
\color{red}\exists\Delta\,B^{j}(l)^{\uparrow\Delta}\, \wedge\forall\delta \leq \Delta\, \Big(I_G(l)^{\uparrow\delta}\,\wedge 	}^{\circled{6}}
\\
& \color{red}\forall\delta' \leq \delta\,
\quad\neg\bigvee_{\substack{l \xrightarrow{\sigma_f, g, r} l'\\ \sigma_f\in \SigmaFor}}(g^{\uparrow\delta'} \wedge I_G(l')^{\uparrow\delta'}[r]\,\wedge
\neg B^{j}(l')^{\uparrow\delta'}[r]) \Big)
	\end{align*}}
	\EndFor
\label{line:BS-update}
\State{$j:=j+1$}
\Until {$\forall l\in L~ B^{j}(l)=B^{j-1}(l)$}
\label{line:BS-end}
\For {$l\in L$}
{$B(l):=B^j(l)$}
\EndFor
\end{algorithmic}
\end{algorithm}

The differences (indicated in red) between Algorithm~\ref{algo:BS} and the bad state condition of EFA presented by~\cite{ouedraogo2011nonblocking} are as follows; 1. The invariant of the target location is considered to determine if the uncontrollable transition should exist in the semantic graph. 2. The third term takes into account the non-preemptable time transitions leading to a bad state.

\begin{property}[\BSP Termination]
\label{prop:BSPtermination}
Given a plant $G$ with the set of locations $L$, set of regions $R_G$, and $\NBP(G)$; Algorithm~\ref{algo:BS} terminates. 
\end{property}

\begin{proof}
See Appendix~\ref{proof:BSPtermination}.
\hfill $\blacksquare$
\end{proof}

\begin{property}[$\BSP$ and Bad States]
\label{prop:BSP}
Given a plant $G$ and $\NBP(G)$: for any $(l,u)$ in (the semantic graph of) $G$, $(l,u)$ is a bad state iff $u\models B(l)$, where $B=\BSP(G,\NBP(G))$.
\end{property}

\begin{proof}
See Appendix~\ref{proof:BSP}.
\hfill $\blacksquare$
\end{proof}

\begin{example}[Bad State Predicate for Bus-Pedestrian]
	\label{ex:BP-BS}
By applying Algorithm~\ref{algo:BS} on the bus-pedestrian example, the bad state predicate of locations $(a,r,0)$ and $(a,c,1)$ are obtained as in Table~\ref{table:BP-BS}. The bad state predicates for $(g,r,\bot)$ and $(g,c,2)$ are $\mathit{true}$ and $\textit{false}$, respectively.

\begin{table}[h]
	\centering
	\captionsetup{width=\columnwidth}
	\caption{Bad state predicate for bus-pedestrian.}
	\label{table:BP-BS}
	\begin{tabular}{|c||c|c|}
		\hline
		 & \multicolumn{2}{c|}{\textbf{$B$}} \\ \hline

		\textbf{$j$} & \textbf{Loc $(a,r,0)$} & \textbf{Loc $(a,c,1)$} \\ \hline
		0 & $x > 2 \vee x-y>1$ & $x > 2$ \\ \hline
		1 & $x \geq 2 \vee x-y>1$  & $ x > 2$ \\ \hline \hline
		2 & $x \geq 2 \vee x-y>1$  & $ x > 2$ \\ \hline
	\end{tabular}
\end{table}
\end{example}

\subsection{Synthesis}
\label{subsection:synthesis}
\begin{figure}[h]
	\centering
	\includegraphics[width=\linewidth]{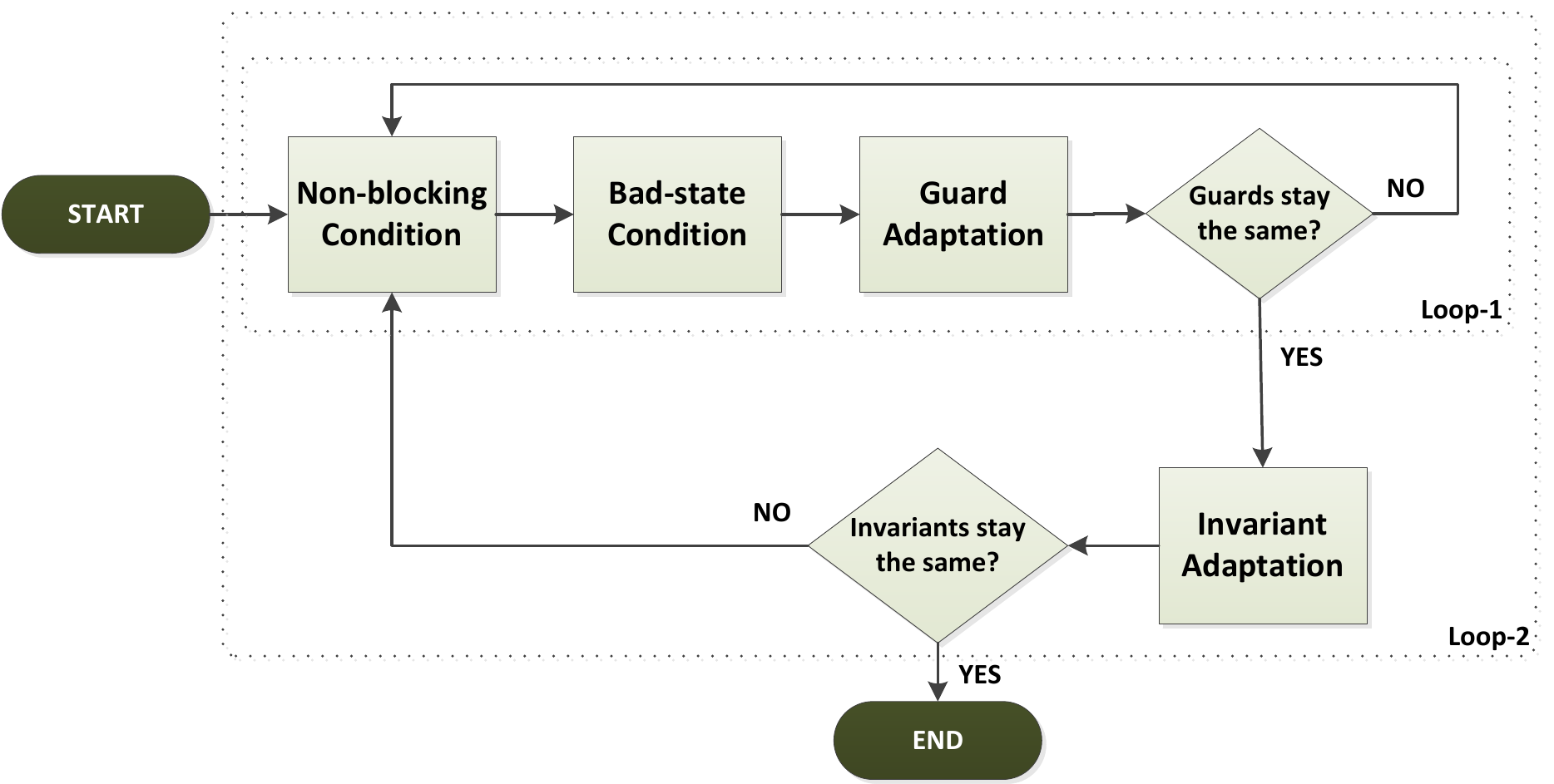}
	\caption{An overview of the synthesis procedure~\cite{Rashidinejad:20}.}
	\label{fig:FlowChart}
\end{figure}

Figure~\ref{fig:FlowChart} gives an overview of the synthesis procedure. As indicated in the figure, there are two loops: 1.\ guard adaptation (Loop-1) considers how the supervisor can affect the controllable events, and 2.\ invariant adaptation (Loop-2) considers how the invariants can be modified using the concept of forcible events.


\subsubsection{Guard adaptation}
Consider Figure~\ref{fig:FlowChart}, in Loop-1 the guards are adapted to obtain a supervisor that prevents the bad states. For this purpose, the guard of each edge $(l,\sigma,g,r,l')$ labeled by a controllable event $\sigma\in\Sigma_c$ is adjusted to become $(l,\sigma,g\wedge \neg B(l')[r],r,l')$.

\subsubsection{Invariant adaptation}
So far, forcible events have not been taken into account. The effect of forcible events preempting time events is taken into account in the invariant adaptation (Loop-2). 
The invariant of a location $l\in L$ can be changed only if there exists an edge labeled by a forcible event $\sigma_f\in \SigmaFor$ starting from $l$. In this case, the invariant is adapted to prevent reaching the bad states as follows:

\begin{equation*}
I(l):=I(l) \wedge \neg{B(l)}.
\end{equation*}


\subsubsection{Synthesis Algorithm}
Algorithm~\ref{algo:synthesis} is the synthesis algorithm.  
For a TA $G$ with a set of uncontrollable events $\Sigmauc$, and a set of forcible events $\Sigma_\mathit{for}$, it results in $S=(C,L,\Sigma_G,E_S,L_m,l_0,I_S)$.
The notation $F_S (l)=\{e\in E_S\mid e.l_s=l,\, e.\sigma\in\SigmaFor,\,e.g \mbox{~is~satisfiable} \}$ gives the set of edges of $S$ starting from location $l$ and labeled by a forcible event.
The algorithm starts with $S=G$.
As indicated in Figure~\ref{fig:FlowChart}, in the inner loop (lines~\ref{line:loop1-s}-\ref{line:loop1-e}), the guards of edges labeled by controllable events are adapted until a fix-point is reached. In the outer loop (lines~\ref{line:loop2-s}-\ref{line:loop2-e}), the invariants of locations where there exist an edge labeled by a forcible event are adapted until a fix-point is reached. Otherwise, the synthesis goes back to Loop-1 (guard adaptation). 
Note that if the invariant of a location $l$ is adapted, and in some later iteration the guard of an edge labeled by the forcible event becomes false, then the invariant should be set back to its original $I_G(l)$. This is captured in line~\ref{line:f-1}.


\renewcommand{\algorithmicrequire}{\textbf{Input: }}
\renewcommand{\algorithmicensure}{\textbf{Output: }}
\begin{algorithm}[h]
\caption{Timed supervisory control synthesis (\TSCS)}
{\algorithmicrequire} {$G=(C,L,\Sigma_G,E_G,L_m,l_0,I_G)$, $\Sigmauc$, $\Sigma_c$, $\Sigma_\mathit{for}$}

{\algorithmicensure} {$S=(C,L,\Sigma_G,E_S,L_m,l_0,I_S)$}
\label{algo:synthesis}
\begin{algorithmic}[1]
\State {$S := G$}
\State{$n:=0$} 
\For {$e\in E_S$, $e=(l,\sigma,g,r,l')$}
{$e.g^0:=e.g$}
\EndFor
\For {$l\in L$}
{$I_S^0(l):=I_G(l)$}
\EndFor
\Repeat\label{line:loop2-s} \Comment{\textcolor{red}{Loop-2: Invariant Adaptation}}
\State {$m:=0$}
\Repeat\label{line:loop1-s} \Comment{\textcolor{red}{Loop-1: Guard Adaptation}}
\State{$N^{n,m}:=\NBP(S)$}
\State{$B^{n,m}:=\BSP(S,N^{n,m})$}
\For{$e\in E_S$ such that $e.\sigma\in\Sigma_c$}
\State{$e.g^{m+1} := e.g^m \wedge \neg B^{n,m}(l')[r]$}\label{line:g-adapt}
\EndFor
\State{$m:=m+1$}
\Until{$\forall {e\in E_S}~ e.g^m= e.g^{m-1}$}\label{line:loop1-e}
\For {$e\in E_S$}
{$e.g:=e.g^m$}
\EndFor
\For{$l\in L$}
\If {$F_S(l)\neq \varnothing$}
{$I^{n+1}_S(l):=I^{n}_S(l) \wedge\neg B^{n,m}(l)$}\label{line:invariant}
\Else { $I^{n+1}_S(l):= I_G(l)$}\label{line:f-1}
\EndIf
\EndFor
\State{$n:=n+1$}
\Until{$\forall{l\in L}~ I^{n}_S(l)= I^{n-1}_S(l)$}\label{line:loop2-e}
\For{$l\in L$}
{$I_S(l):=I^n_S(l)$}
\EndFor
\end{algorithmic}
\end{algorithm}

Given a plant $G$, in case that $u_0\models B(l_0)$, with $B$ as the result of Algorithm~\ref{algo:BS} for $\TSCS(G)$ and $\NBP(\TSCS(G))$, then $\TSCS(G)$ is undefined. In the rest of the paper, it is assumed that $u_0\not\models B(l_0)$ for any given plant $G$.



\begin{property}[\TSCS Termination]
\label{prop:TSCS termination}
Given a plant $G$; Algorithm~\ref{algo:synthesis} terminates.
\end{property}


\begin{proof}
See Appendix~\ref{proof:TSCS termination}.
\hfill $\blacksquare$
\end{proof}

\begin{property}[$\TSCS(G)$ is a TA]
\label{prop:S is TA}
Given a plant $G$, $S=\TSCS(G)$ is a TA.
\end{property}

\begin{proof}
See Appendix~\ref{proof:S is a TA}.
\hfill $\blacksquare$
\end{proof}

\begin{property}[$\TSCS(G)$ is a subautomaton of $G$]
\label{property:subG}
Given a plant $G$: $\TSCS(G)\subseteq G$.
\end{property}

\begin{proof}
See Appendix~\ref{proof:subG}.
\hfill $\blacksquare$
\end{proof}

According to Property~\ref{property:subG}, $\TSCS(G)||G =\TSCS(G)$. 

\begin{property}[Algorithm Correctness]
\label{property:base4proofs}
Given a plant $G$ and the supervisor $S=\TSCS(G)$:
for any reachable state $(l,u)$ (in the semantic graph) of $S$: $u\not\models B(l)$, where $B=\BSP(S,\NBP(S))$.
\end{property}

\begin{proof}
See Appendix~\ref{proof:base4proofs}.
\hfill $\blacksquare$
\end{proof}

The following theorems summarize the main results of the paper.

\begin{theorem}[Controllability]
\label{theorem:controllability}
Given a plant $G$ with uncontrollable events $\Sigmauc$ and forcible events $\SigmaFor$, and the supervisor $S=\TSCS(G)$: $S$ is controllable w.r.t.\ $G$.
\end{theorem}

\begin{proof}
See Appendix~\ref{proof:controllability}.
\hfill $\blacksquare$
\end{proof}

\begin{theorem}[Nonblockingness]
\label{theorem:NBness}
Given a plant $G$ and the supervisor $S=\TSCS(G)$: the supervised plant $S||G$ is nonblocking.
\end{theorem}

\begin{proof}
See Appendix~\ref{proof:NBness}.
\hfill $\blacksquare$
\end{proof}

\begin{theorem}[Maximal Permissiveness]
\label{theorem:MPness}
Given a plant $G$ and the supervisor $S=\TSCS(G)$: $S$ is maximally permissive for $G$.
\end{theorem}

\begin{proof}
See Appendix~\ref{proof:MPness}.
\hfill $\blacksquare$
\end{proof}

\begin{example}[Supervisor Synthesis for Bus-Pedestrian]
\label{ex:BP-synthesis}
Let us apply Algorithm~\ref{algo:synthesis} to the bus-pedestrian from Example~\ref{ex:busped}.
Initially, $S$ is set to the plant depicted in Figure~\ref{fig:BP-synch}.
First, the guard of the edge labeled by the controllable event $jump$ is modified to $y\geq 1\wedge x\leq 2$. Since $N^{1,0}=N^{0,0}$ and also $B^{1,0}=B^{0,0}$, $e.g^1=e.g^0$, and the inner loop stops. 
Next, for $l_0=(a,r,0)$, the invariant is adapted to $x\leq 2\wedge x<2=x<2$.
Since $N^{1,1}=N^{1,0}$ and also $B^{1,1}=B^{1,0}$, $I^{1}_S(l_0)=I^{0}_S(l_0)$ and the outer loop also terminates. The synthesized supervisor is depicted in Figure~\ref{fig:BP-S}.

\begin{figure}[h]
\centering
\begin{tikzpicture}[>=stealth', shorten >=1pt, auto, node distance=3.5cm, scale=.75, 
transform shape, align=center,state/.style={circle, draw, minimum size=1.5cm,font=\small}]

    \node[initial,initial text={},state]           (A)                                    {$(a,r,0)$ \\ \textcolor{red}{$x< 2 \wedge$} \\ \textcolor{red}{$x-y\leq 1$}};
    \node[state]         (B) [right of=A]                       {$(g,r,\bot)$};
    \node[state]         (C) [below of=A]                     {$(a,c,1)$ \\ $x\leq 2$};
    \node[state,accepting]         (D) [right of=C]                       {$(g,c,2)$};

    \path[->] (A) edge [above,dashed]   node [align=center]  {\small{$x=2$}\\ \small{$pass$}} (B)
    (A) edge [right]   node [align=center]  {\small{$y\geq 1 \wedge {}$}  \textcolor{red}{\small{$x\leq 2$}}\\ \small{\underline{$jump$}}} (C)
    (C) edge [above,dashed]   node [align=center]  {\small{$x=2$}\\ \small{$pass$}} (D);
\end{tikzpicture}
\caption{Supervisor for bus-pedestrian from Example~\ref{ex:busped}.}
\label{fig:BP-S}
\end{figure}
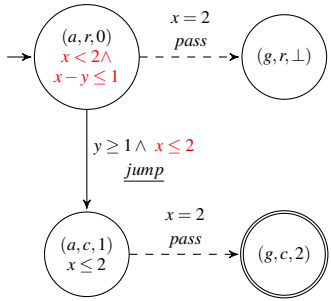
\end{example}

\begin{remark}
Invariant adaptation can highly affect the  synthesis result. Consider Example~\ref{ex:BP-synthesis}, Algorithm~\ref{algo:synthesis} does not result in a supervisor without invariant adaptation.
However, if the TA has no forcible event, time transitions are always uncontrollable and the synthesis procedure can be adjusted as follows: 
\begin{enumerate}
    \item the update of the bad state predicate (Algorithm~\ref{algo:BS}-line~\ref{line:BS-update}) simplifies to
\begin{align*}
& B^{j+1}(l) := ~B^{j}(l) \vee \bigvee_{\substack{l \xrightarrow{\sigma, g, r} l'\\ \sigma\in \Sigmauc}} \big(g  \wedge I_G(l')[r]\wedge B^{j}(l')[r]\big)
\,\vee\\
&~\exists\Delta\; B^{j}(l)^{\uparrow\Delta}\, \wedge\forall\delta \leq \Delta\; I_G(l)^{\uparrow\delta}
\end{align*}
where the last part of \circled{6} is removed, and
\item the algorithm ends after the inner loop indicated in Figure~\ref{fig:FlowChart} since guard adaptation is the only modification that can be applied through synthesis.
\end{enumerate}
\end{remark}

\section{Requirement Automata}
\label{section:requirement}
To generalize the method to a wider class of applications, we solve the TSC synthesis problem for a given set of control requirements. 
It is assumed that an allowed behavior of $G$ is denoted by the timed automaton $R=(C_R,Q,\Sigma_R,E_R,Q_m,q_0,I_R)$, where $\Sigma_R\subseteq \Sigma_G$ and $C_R\cap C=\emptyset$. 
Since most control requirements are defined to provide safety of a plant, we call a supervised plant $SP=S||G$ \emph{safe} if it satisfies the control requirement $R$.

\begin{defn}[Safety]
\label{dfn:safety}
Given a plant $G$ and a control requirement $R$, a TA $S$ with event set $\Sigma_S$ is safe w.r.t.\ $G$ and $R$ if $P_{\Sigma_{\mathit{SP}}\cap\Sigma_R}(L({\mathit{S||G}}))\subseteq P_{\Sigma_{\mathit{SP}}\cap\Sigma_R}(L(R))$ with $\Sigma_{\mathit{SP}}=\Sigma_S\cup\Sigma_G$. 
\hfill $\blacksquare$
\end{defn}

Requirement automata can be considered in synthesis by being transferred into the plant using synchronous product. However, if a requirement automaton is not controllable (Definition~\ref{dfn:controllability}), then it is necessary to let the supervisor know about the uncontrollable events that are disabled by a given requirement.
To take care of this issue, a requirement automaton $R$ is made complete.
\emph{Completion} was first introduced in~\cite{Flordal:07} for DES, where the requirement automaton $R$ is made complete as $R^{\bot}$ in terms of uncontrollable events.
By applying the synthesis on $G||R^\bot$, all original controllability problems in $G||R$ are translated to blocking issues.
To solve the blocking issues, synthesis still takes the controllability definition into account.
Inspired from~\cite{Flordal:07}, we present the completion of a TA.

\begin{defn}[TA Completion]
\label{dfn:Rtot}
Given a TA $R=(C_R,Q,\Sigma,E_R,Q_m,q_0,I_R)$, the complete automaton $R^{\bot}$ is defined as
$R^{\bot}=(C_R,Q\cup\{q_d\},\Sigma,E^{\bot}_R,Q_m,q_0,I_R)$, where $q_d\notin Q$, $I_R(q_d)=\true$ and $I_R(q) = I(q)$ for all $q \in Q$, and for every $q_{s}\in Q$, $\sigma\in \Sigmauc$:
\begin{equation*}
E^{\bot}_R=E_R\cup \{ (q_{s}, \sigma, g^{\bot}, \{\}, q_d) \mid (q_s,\sigma,g,r,q_t) \in E_R \},
\end{equation*}
where $g^{\bot} = \neg \big(\bigvee_{e \in E_R, e.q_s = q_s, e.\sigma=\sigma} e.g \land I_R(e.q_t)[e.r]\big)$.
\hfill$\blacksquare$
\end{defn}

To synthesize a supervisor, Algorithm~\ref{algo:synthesis} is applied on $G||R^\bot$.
The obtained supervisor is already guaranteed to be controllable, maximally permissive, and it results in a nonblocking supervised plant.
Theorem~\ref{theorem:safety} shows that the supervised plant is safe as well.

\begin{theorem}[Safety]
\label{theorem:safety}
Given a plant $G$, a set of control requirements $R$, and the supervisor $S=\TSCS(G||R^\bot)$: $S$ is safe for $G$ w.r.t.\ $R$.
\end{theorem}

\begin{proof}
See Appendix~\ref{proof:safety}.
\hfill$\blacksquare$
\end{proof}

In general, there can be a set of control requirements $\{R_1,R_2,\ldots,R_n\}$ given for a plant.
In that case, the allowed behavior of $G$, is determined by the synchronous product of all requirement automata; $R=R_1||R_2||\ldots||R_n$.
Since completion distributes over synchronous product,
$R^\bot$ can be computed either as $R^\bot_1||R^\bot_2||\ldots||R^\bot_n$, or $(R_1||R_2||\ldots||R_n)^\bot$.

\section{Case Study}
\label{section:CaseStudy}
\begin{figure}[h]
    \centering
    \begin{tikzpicture}[>=stealth', shorten >=1pt, auto, node distance=3.2cm, scale=.75, transform shape, align=center,state/.style={circle, draw, minimum size=1.2cm,font=\small}]

    \node[initial above,initial text={},state,accepting]           (A)                                    {$c_0$};
    \node[state]         (B) [right of=A]                       {$c_1$\\$z\leq 1$};
    \node[state]         (C) [left of=A]                       {$c_2$\\$z\leq 1$};

    \path[->] (A) edge [above,dashed]   node [align=center]  {\small{}\\ \small{$\app$ \quad $z:=0$}} (B)
    (B) edge [bend left]   node [align=center]  {\small{$z=1$}\\ \small{$\underline{\lowerE}$}} (A)
    (A) edge [above]   node [align=center]  {\small{}\\ \small{$z:=0$ \quad $\exit$}} (C)
    (C) edge [bend right]   node [align=center]  {\small{}\\ \small{$\underline{\raiseE}$}} (A)
    ;
    \end{tikzpicture}
    \caption{Gate-controller from~\cite{alur1994theory,alur1999timed}.}
    \label{fig:controller}
\end{figure}
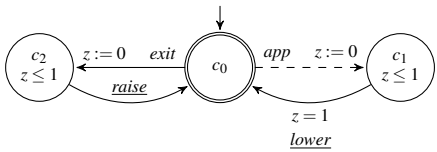

In this section, we consider the verification example from~\cite{alur1994theory,alur1999timed} and modify it for  synthesis. The TA representing the train and gate are depicted in Figure~\ref{fig:TG}. The system in~\cite{alur1994theory,alur1999timed} also involves an automatic controller, depicted in Figure~\ref{fig:controller}, to open and close the gate in a railroad crossing.  The control requirements for the train-gate-controller system are as follows~\cite{alur1994theory}:

\begin{itemize}
    \item Safety requirement: whenever the train is inside the gate, the gate should be closed.
    \item Liveness requirement: the gate is never closed for more than 10 time units.
\end{itemize}

In \cite{alur1994theory,alur1999timed}, the system is assessed to be safe by analysing the timing constraints: they say that with the (random) gate-controller, that is part of the system, the event $\lowerE$ always proceeds the event $\inE$, so the system is always safe. We do not consider such a controller to already be given as a part of the system. We synthesize a supervisor that is correct-by-construction, and more importantly this supervisor guarantees  controllability, nonblockingness, and maximal permissiveness.

The models of train and gate are taken directly from~\cite{alur1994theory,alur1999timed}. 
The events $\app$ and $\out$ for the train, and the events $\down$ and $\up$ for the gate are assumed to be uncontrollable. Moreover, the events $\raiseE$ and $\lowerE$ of the gate are assumed to be forcible.  

\begin{figure}[h]
    \centering
    \begin{subfigure}[b]{0.4\columnwidth}
        \centering
        \begin{tikzpicture}[>=stealth', shorten >=1pt, auto, node distance=3.2cm, scale=.7, transform shape, align=center,state/.style={circle, draw, minimum size=1.2cm,font=\small}]

    \node[initial,initial text={},state,accepting]           (A)                                    {$t_0$};
    \node[state]         (B) [right of=A]                       {$t_1$\\$x\leq 5$};
    \node[state]         (C) [below of=B]                       {$t_2$\\$x\leq 5$};
    \node[state]         (D) [below of=A]                       {$t_3$\\$x\leq 5$};

    \path[->] (A) edge [above,dashed]   node [align=center]  {\small{}\\ \small{$\app$}\quad \small{$x:=0$}} (B)
    (B) edge [right]   node [align=center]  {\small{$x>2$}\\ \small{$\inE$}} (C)
    (C) edge [above,dashed]   node [align=center]  {\small{}\\ \small{$\out$}} (D)
    (D) edge [right]   node [align=center]  {\small{}\\ \small{$\exit$}} (A)
    ;
    \end{tikzpicture}
    \caption{Train}
    \end{subfigure}
\hspace{0.4cm}
    \begin{subfigure}[b]{0.4\columnwidth}
        \centering
        \begin{tikzpicture}[>=stealth', shorten >=1pt, auto, node distance=3.2cm, scale=.7, transform shape, align=center,state/.style={circle, draw, minimum size=1.2cm,font=\small}]

    \node[initial,initial text={},state,accepting]           (A)                                    {$g_0$};
    \node[state]         (B) [right of=A]                       {$g_1$\\$y\leq 1$};
    \node[state]         (C) [below of=B]                       {$g_2$};
    \node[state]         (D) [below of=A]                       {$g_3$\\$y\leq 2$};

    \path[->] (A) edge [above]   node [align=center]  {\small{}\\ \small{$\underline{\lowerE}$}\quad \small{ $y:=0$}} (B)
    (B) edge [right,dashed]   node [align=center]  {\small{}\\ \small{$\down$}} (C)
    (C) edge [above]   node [align=center]  {\small{}\\  \small{$y:=0$}\quad \small{$\underline{\raiseE}$}} (D)
    (D) edge [right,dashed]   node [align=center]  {\small{$y\geq 1$}\\ \small{$\up$}} (A)
    ;
    \end{tikzpicture}
    \caption{Gate}
    \end{subfigure}
\caption{Train-gate system.}
\label{fig:TG}
\end{figure}
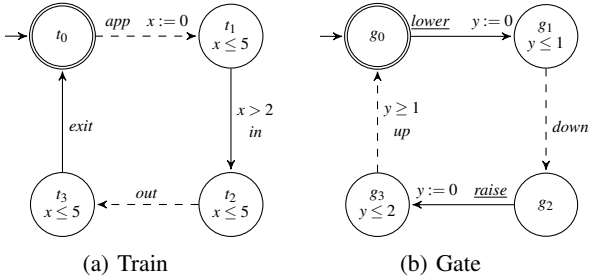

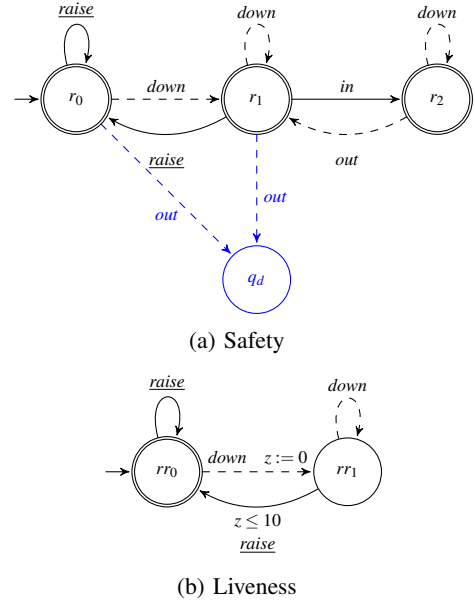
\begin{figure}[h]
    \centering
    \begin{subfigure}[b]{0.8\columnwidth}
        \centering
        \begin{tikzpicture}[>=stealth', shorten >=1pt, auto, node distance=3.2cm, scale=.75, transform shape, align=center,state/.style={circle, draw, minimum size=1.2cm,font=\small}]

    \node[initial,initial text={},state,accepting]           (A)                                    {$r_0$};
    \node[state,accepting]         (B) [right of=A]                       {$r_1$};
    \node[state,accepting]         (C) [right of=B]                       {$r_2$};
    \node[state,blue]         (D) [below of=B]                       {$q_d$};
    
    \path[->] (A) edge [above,dashed]   node [align=center]  {\small{}\\ \small{$\down$}} (B)
    (B) edge [bend left]   node [align=center]  {\small{}\\ \small{$\underline{\raiseE}$}} (A)
    (A) edge [loop above]   node [align=center]  {\small{}\\ \small{$\underline{\raiseE}$}} (A)
    (B) edge [loop above,dashed]   node [align=center]  {\small{}\\ \small{$\down$}} (B)
    (B) edge [above]   node [align=center]  {\small{}\\ \small{$\inE$}} (C)
    (C) edge [bend left,dashed]   node [align=center]  {\small{}\\ \small{$\out$}} (B)
    (C) edge [loop above,dashed]   node [align=center]  {\small{}\\ \small{$\down$}} (C)
    
    (A) edge [below,dashed,blue]   node [align=center]  {\small{}\\ \small{$\out$}} (D)
    (B) edge [dashed,blue]   node [align=center]  {\small{}\\ \small{$\out$}} (D)
    ;
    \end{tikzpicture}
    \caption{Safety}
    \label{fig:safety}
    \end{subfigure}
\hspace{0.4cm}
    \begin{subfigure}[b]{0.5\columnwidth}
        \centering
        \begin{tikzpicture}[>=stealth', shorten >=1pt, auto, node distance=3.2cm, scale=.75, transform shape, align=center,state/.style={circle, draw, minimum size=1.2cm,font=\small}]

    \node[initial,initial text={},state,accepting]           (A)                                    {$rr_0$};
    \node[state]         (B) [right of=A]                       {$rr_1$};

    \path[->] (A) edge [above,dashed]   node [align=center]  {\small{$\down$}\quad \small{$z:=0$}} (B)
    (B) edge [bend left]   node [align=center]  {\small{$z\leq 10$}\\ \small{$\underline{\raiseE}$}} (A)
    (A) edge [loop above]   node [align=center]  {\small{}\\ \small{$\underline{\raiseE}$}} (A)
    (B) edge [loop above,dashed]   node [align=center]  {\small{}\\ \small{$\down$}} (B);
    \end{tikzpicture}
    \caption{Liveness}
    \label{fig:liveness}
    \end{subfigure}
\caption{Requirements for train-gate system.}
\label{fig:TG-Rs}
\end{figure}

The safety requirement is represented by the TA in Figure~\ref{fig:safety},  where the blue  location and edges are added to make the TA complete. 
The liveness requirement is represented by the TA in Figure~\ref{fig:liveness}. The liveness requirement does not need completion as the uncontrollable event $\down$ is enabled at both states of the automaton. 

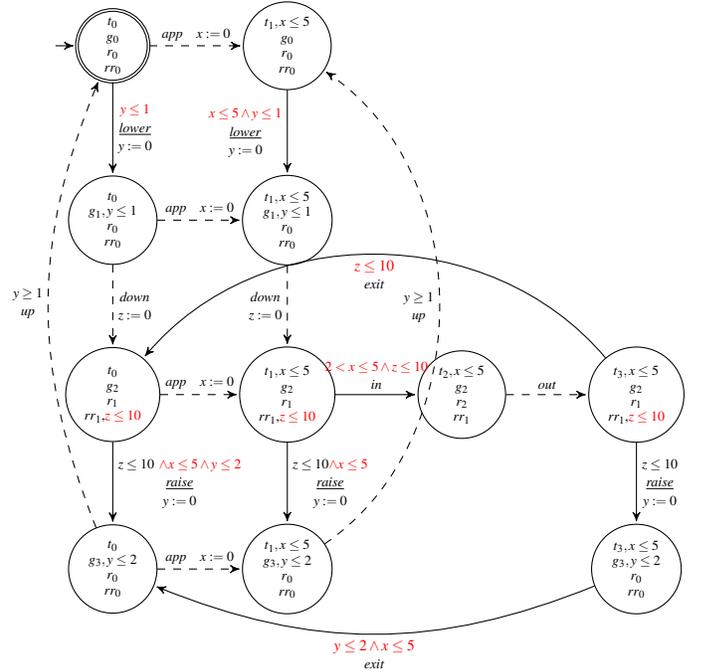
\begin{figure}[h]
\centering
    \begin{tikzpicture}[>=stealth', shorten >=1pt, auto, node distance=4cm, scale=.58, transform shape, align=center,state/.style={circle, draw, minimum size=0.9cm,font=\small}]

    \node[initial,initial text={},state,accepting]           (p0)                                    {$t_0$\\
    $g_0$\\$r_0$\\$rr_0$};
    \node[state]         (p1) [right of=p0]                       {$t_1,x\leq 5$\\$g_0$\\$r_0$\\$rr_0$};
    \node[state]         (p2) [below of=p0]                       {$t_0$\\$g_1,y\leq 1$\\$r_0$\\$rr_0$};
    \node[state]         (p3) [right of=p2]                       {$t_1,x\leq 5$\\$g_1,y\leq 1$\\$r_0$\\$rr_0$};
    
    \node[state]         (p4) [below of=p2]                       {$t_0$\\$g_2$\\$r_1$\\$rr_1$,\small\textcolor{red}{$z\leq 10$}};
    \node[state]         (p5) [right of=p4]                       {$t_1,x\leq 5$\\$g_2$\\$r_1$\\$rr_1$,\small\textcolor{red}{$z\leq 10$}};
    \node[state]         (p6) [right of=p5]                       {$t_2,x\leq 5$\\$g_2$\\$r_2$\\$rr_1$};
    \node[state]         (p7) [right of=p6]                       {$t_3,x\leq 5$\\$g_2$\\$r_1$\\$rr_1$,\small\textcolor{red}{$z\leq 10$}};

    \node[state]         (p8) [below of=p4]                       {$t_0$\\$g_3,y\leq 2$\\$r_0$\\$rr_0$};
    \node[state]         (p9) [right of=p8]                       {$t_1,x\leq 5$\\$g_3,y\leq 2$\\$r_0$\\$rr_0$};
    \node[state]         (p11) [below of=p7]                       {$t_3,x\leq 5$\\$g_3,y\leq 2$\\$r_0$\\$rr_0$};

    \path[->] (p0) edge [above,dashed]   node [align=center]  {\small{$\app$}\quad \small{$x:=0$}} (p1)
    (p2) edge [above,dashed]   node [align=center]  {\small{$\app$}\quad \small{$x:=0$}} (p3)
    (p4) edge [above,dashed]   node [align=center]  {\small{}\\ \small{$\app$}\quad \small{$x:=0$}} (p5)
    (p8) edge [above,dashed]   node [align=center]  {\small{}\\ \small{$\app$}\quad \small{$x:=0$}} (p9)

    (p0) edge [right]   node [align=center]  {\small\textcolor{red}{$ y\leq 1$}\\ \small{$\underline{\lowerE}$}\\ \small{$y:=0$}} (p2)
    (p1) edge [left]   node [align=center]  {\small\textcolor{red}{$x\leq 5 \wedge y\leq 1$}\\ \small{$\underline{\lowerE}$}\\ \small{$y:=0$}} (p3) 

    (p2) edge [right,dashed]   node [align=center]  {\small{$\down$}\\ \small{$z:=0$}} (p4)
    (p3) edge [left,dashed]   node [align=center]  {\small{$\down$}\\ \small{$z:=0$}} (p5)
    
    (p4) edge [right]   node [align=center]  {\small{$z\leq 10$} \small\textcolor{red}{$\wedge x\leq 5\wedge y\leq 2$}\\ \small{$\underline{\raiseE}$} \\ \small{$y:=0$}} (p8)
    (p5) edge [right]   node [align=center]  {\small{$z\leq 10$}\textcolor{red}{$\wedge x\leq 5$}\\ \small{$\underline{\raiseE}$} \\ \small{$y:=0$}} (p9)
    (p7) edge [right]   node [align=center]  {\small{$z\leq 10$}\\ \small{$\underline{\raiseE}$} \\ \small{$y:=0$}} (p11)

    (p5) edge [above]   node [align=center]  {\small\\\small\textcolor{red}{$2<x\leq 5\wedge z\leq 10$}\\ \small{$\inE$}} (p6)
    (p6) edge [above,dashed]   node [align=center]  {\small{}\\ \small{$\out$}} (p7)

    (p7) edge [bend right=50]   node [align=center]  {\textcolor{red}{$z\leq 10$}\\ \small{$\exit$} \\ \small{}} (p4)
    (p11) edge [bend left=22]   node [align=center]  {\textcolor{red}{$y\leq 2\wedge x\leq 5$}\\ \small{$\exit$} \\ \small{}} (p8)

    (p8) edge [bend left=22,dashed]   node [align=center]  {\small{$y\geq 1$}\\ \small{$\up$}} (p0)
    (p9) edge [bend right=56,dashed]   node [align=center]  {\small{$y\geq 1$}\\ \small{$\up$}} (p1)
    ;
    \end{tikzpicture}
    \caption{Synthesized supervisor for train-gate and control requirements. Edges with guards equal to \false and locations reached by them have been removed.}
    \label{fig:TGRs}
\end{figure}

The supervisor synthesized by Algorithm~\ref{algo:synthesis} for the train-gate and control requirements is given in Figure~\ref{fig:TGRs}. In this figure, the synchronous product of the train-gate and control requirements is indicated in black and the adaptations made by the supervisor in red.

\section{Conclusion and future Work}
\label{section:conclusion}
In this paper, we propose a synthesis algorithm for timed automata (TA) with a set of forcible events. The algorithm is directly applicable on TA without abstracting them to finite state automata.
The objective is to avoid  blocking states. To take care of controllability, not only the blocking states but also the states from which a blocking state is reachable in an uncontrollable manner (referred to as bad states) should be avoided.
The bad states are determined using nonblocking and bad state predicates associated to each location.
The modifications made through synthesis are as follows: 1. guard adaptation of edges labeled by controllable events, and 2. invariant adaptation of locations from which there exist an edge labeled by a forcible event.
Based on the notion of extended clock regions, it is proven that the synthesized supervisor satisfies nonblockingness, controllability, and maximal permissiveness.
To generalize, we solve the problem for a given set of control (safety) requirements modeled as TA. We guarantee that the synthesized supervisor satisfies controllability, nonblockingness, maximal permissiveness, and safety.
Finally, the results are verified by applying the method on a case study.
Networked supervisory control of timed automata will be studied in future research. Moreover, implementation of the proposed approach in available tool sets will be investigated.

\section*{Acknowledgment}
The authors would like to thank Patrick van der Graaf for his initial efforts on this subject.

\appendices
\section{Technical Lemmas}

\begin{lemma}[$G$-Clock Constraint]
\label{lemma:G-cc}
$\varphi$ is a $G$-clock constraint iff for any pair of clock valuations $u_1,u_2$, represented by the same clock region $r_G\in R_G$: $u_1\models \varphi \Longleftrightarrow u_2\models \varphi$.
\end{lemma}


\begin{proof}
The proof is trivial.
%
\hfill$\blacksquare$
\end{proof}

\begin{lemma}[Clock Valuations]
\label{lemma:CV}
For any pair of clock valuations $u_1,u_2\in r_G$ for some $r_G\in R_G$, if $u_1+\Delta_1\in r_{\Delta}$ for some $r_{\Delta}\in R_G$ and $\Delta_1\in\RealTime$: there exists $\Delta_2\in\RealTime$ such that $u_2+\Delta_2\in r_{\Delta}$.
\end{lemma}

\begin{proof}
As illustrated by Figure~\ref{fig:Alur-Example}, from two valuations from the same region in each case any move to another region (by passage of time) from one of these valuations is easily mimicked from the other valuation (possibly for a different amount of time passage).
\hfill$\blacksquare$
\end{proof}

\begin{remark}
In the coming lemmas and proofs, we frequently use ''this term represents a $G$-clock constraint". The meaning of this is that although the term may not necessarily satisfy $G$-clock constraints as given by Definition~\ref{defn:$G$-cc}, there is $G$-clock constraint that is logically equivalent with it (which means that for any valuation the term and its $G$-clock constraint representation have the same value).
\end{remark}

\begin{lemma}[Negation of $G$-Clock Constraint]
\label{lemma:neg}
For any $G$-clock constraint $\varphi$, the negation $\neg \varphi$ also represents a $G$-clock constraint.
\end{lemma}

\begin{proof}
This is proved by induction on the structure of $G$-clock constraints.

\textbf{Base cases:}
\begin{itemize}
    \item for the atomic $G$-clock constraints $x < n$ and $x-y < n$, their negations are $x \geq n$ and $x-y \geq n$, respectively;
    \item for the atomic $G$-clock constraints $x = n$ and $x-y = n$, their negations are $x > n \vee x < n$ and $x-y > n \vee x-y < n$, respectively;
    \item for the atomic $G$-clock constraints $x > n$ and $x-y > n$, their negations are $x \leq n$ and $x-y \leq n$, respectively.
\end{itemize}

\textbf{Induction step:} Consider the $G$-clock constraint $\varphi=\varphi_1\Diamond\varphi_2$ for some $G$-clock constraints $\varphi_1$ and $\varphi_2$, and $\Diamond \in \{ \wedge, \vee \}$, where the statement holds for $\varphi_1$ and $\varphi_2$, i.e., $\neg\varphi_1$ and $\neg\varphi_2$ also represent $G$-clock constraints. 
If $\Diamond=\vee$, then
$\neg(\varphi_1 \Diamond \varphi_2) = \neg\varphi_1 \wedge \neg\varphi_2$.  Also, if $\Diamond=\wedge$, then $\neg(\varphi_1 \Diamond \varphi_2) = \neg\varphi_1 \vee \neg\varphi_2$. Both $\neg\varphi_1$ and $\neg\varphi_2$ represent $G$-clock constraints as assumed, and the combination of any two $G$-clock constraints by $\wedge$ and $\vee$ is also a $G$-clock constraint. So, $\neg(\varphi_1 \Diamond \varphi_2)$ represents a $G$-clock constraint.

\textbf{Conclusion:} By the principle of induction, the claim of Lemma~\ref{lemma:neg} holds for any $G$-clock constraint $\varphi$.
\hfill$\blacksquare$
\end{proof}

\begin{lemma}[Reset Update of $G$-Clock Constraint]
\label{lemma:reset}
For any $G$-clock constraint $\varphi$ and any reset $r$, $\varphi[r]$ also represents a $G$-clock constraint.
\end{lemma}

\begin{proof}
This is proved by induction on the structure of $G$-clock constraints.


\textbf{Bases cases: }

\begin{itemize}
    \item for the atomic $G$-clock constraints $x<n$ and $x-y<n$, $\varphi[r]$ represents the $G$-clock constraint $\varphi$ if $x,y\notin r$. If $x,y\in r$, $\varphi[r]$ represents the $G$-clock constraint $\true$ if $n\neq 0$ and $\false$ if $n=0$. For $x-y<n$, $\varphi[r]$ represents the $G$-clock constraint $y>n$ if only $x\in r$, and $x<n$ if only $y\in r$. 

    \item for the atomic $G$-clock constraints $x=n$ and $x-y=n$, $\varphi[r]$ represents the $G$-clock constraint $\varphi$ if $x,y\notin r$. If $x,y\in r$, $\varphi[r]$ represents the $G$-clock constraint $\true$ if $n=0$ and $\false$ if $n\neq 0$. For $x-y=n$, $\varphi[r]$ represents the $G$-clock constraint $y=n$ if only $x\in r$, and $x=n$ if only $y\in r$.

    \item for the atomic $G$-clock constraints $x>n$ and $x-y>n$, $\varphi[r]$ represents the $G$-clock constraint $\varphi$ if $x,y\notin r$. If $x,y\in r$, $\varphi[r]$ represents the $G$-clock constraint $\false$. For $x-y>n$, $\varphi[r]$ represents the $G$-clock constraint $y<n$ if only $x\in r$, and $x>n$ if only $y\in r$.
\end{itemize}

\textbf{Induction step:} Consider the $G$-clock constraint $\varphi = \varphi_1 \Diamond \varphi_2$ for some $G$-clock constraints $\varphi_1$ and $\varphi_2$, and $\Diamond \in \{ \wedge, \vee \}$, where the statement holds for $\varphi_1$ and $\varphi_2$, i.e., $\varphi_1[r]$ and $\varphi_2[r]$ also represent $G$-clock constraints.
$(\varphi_1 \Diamond \varphi_2)[r] = \varphi_1[r] \Diamond \varphi_2[r]$ because the reset update does not change anything else than replacing all clock variables of $r$ by zero. So, in $(\varphi_1\Diamond \varphi_2)[r]$, the clock variables from $r$ in both $\varphi_1$ and $\varphi_2$ are replaced by zero which can equivalently be represented by $\varphi_1[r]\Diamond \varphi_2[r]$.
Since the combination of any two $G$-clock constraints by $\wedge$ and $\vee$ is also a $G$-clock constraint, $(\varphi_1 \Diamond \varphi_2)[r]$ represents a $G$-clock constraint.

\textbf{Conclusion:} By the principle of induction, the claim of Lemma~\ref{lemma:reset} holds for any $G$-clock constraint $\varphi$.
\hfill$\blacksquare$
\end{proof}





\begin{lemma}[$\Delta$-Time Invariance for $\NBP$]
\label{lemma:Delta-NBP}
Given $G$-clock constraints $\varphi_1$ and $\varphi_2$,  $\exists\Delta\; \varphi_1^{\uparrow\Delta}\wedge$ $\forall\delta\leq \Delta\; \varphi_2^{\uparrow\delta}$ represents a $G$-clock constraint.
\end{lemma}

\begin{proof}
Let us indicate $\exists\Delta\; \varphi_1^{\uparrow\Delta}\wedge$ $\forall\delta\leq \Delta\; \varphi_2^{\uparrow\delta}$ by $\Phi$.
Take a clock valuation $u_1$ represented by a clock region of $G$, say $r_G\in R_G$, such that $u_1\models \Phi$.
According to Lemma~\ref{lemma:G-cc}, it suffices to prove that for any region $r_G$ and any two clock valuations $u_1$ and $u_2$ represented by $r_G$, $u_1 \models \Phi$ iff $u_2\models \Phi$. Because of symmetry considerations it suffices to prove that $u_1 \models \Phi$ implies $u_2 \models \Phi$.
Let us assume $u_1\models \Phi$. Then there exists some $\Delta_1$ such that $u_1\models \varphi_1^{\uparrow\Delta_1}$ and  $u_1\models\forall\delta\leq \Delta_1\; \varphi_2^{\uparrow\delta}$.
It is proved that there always exists a $\Delta_2$ for which $u_2\models \varphi_1^{\uparrow\Delta_2}$ and $u_2\models\forall\delta\leq \Delta_2\; \varphi_2^{\uparrow\delta}$.
Let us say $u_1^{\uparrow\Delta_1}\in r_{\Delta}$ for some $r_{\Delta}\in R_G$. Then, based on Lemma~\ref{lemma:CV}, there exists a $\Delta_2\in\RealTime$ such that $u_2^{\uparrow\Delta_2}\in r_{\Delta}$.
Since $u_1^{\uparrow{\Delta_1}}\models \varphi_1$ and $u_1^{\uparrow{\Delta_1}},u_2^{\uparrow{\Delta_2}}\in r_{\Delta}$, by Lemma~\ref{lemma:G-cc}: $u_2^{\uparrow{\Delta_2}}
\models \varphi_1$.
It suffices to prove that for all $\delta\leq \Delta_2$: $u_2\models \varphi_2^{\uparrow{\delta}}$.
Take $\delta_2\leq \Delta_2$, and assume that $u_2^{\uparrow{\delta_2}}\in r_{\delta}$, where $r_{\delta}$ can be $r_G$, $r_{\Delta}$, or any region in between.
Based on Lemma~\ref{lemma:CV}, there exists a $\delta_1$ such that $u_1^{\uparrow{\delta_1}}\in r_{\delta}$. We prove that $\delta_1\leq \Delta_1$ by contradiction. Assume $\delta_1>\Delta_1$. Then, $r_{\delta}$ already passed $r_\Delta$, and this contradicts the fact that $r_{\delta}$ is either $r_G$, $r_{\Delta}$, or any other region in between.
So, $\delta_1\leq\Delta_1$, and $u_2^{\uparrow{\delta_2}}\models \varphi_2$ because $u_1^{\uparrow{\delta_1}}\models \varphi_2$, and $u_1^{\uparrow{\delta_1}},u_2^{\uparrow{\delta_2}}\in r_{\delta}$.
\hfill$\blacksquare$
\end{proof}

\begin{lemma}[Nonblocking Predicate]
\label{lemma:NB is cc}
Given a plant $G$, $N^i(l)$ computed by Algorithm~\ref{algo:NB} in each iteration $i$ and for each location $l\in L$ represents a $G$-clock constraint.
\end{lemma}

\begin{proof}
We do the proof by induction on the number of iterations $i$.

\textbf{Base case:} $i=0$. Then, $N^0(l)$ is either $I_G(l)$ or $\false$, and in each case, this is a $G$-clock constraint by definition.

\textbf{Induction step:} Assume that the statement holds for $i$, i.e., $N^i(l)$ is a $G$-clock constraint for all $l\in L$. It suffices to prove that the statement holds for $i+1$, i.e., $N^{i+1}(l)$ is a $G$-clock constraint for all $l \in L$.
Consider Algorithm~\ref{algo:NB}-line~\ref{line:NB-update}, $N^{i+1}=\circled{1}\vee\circled{2}\vee\circled{3}$. It suffices to prove that each of $\circled{1}$, $\circled{2}$, and $\circled{3}$ is a $G$-clock constraint because then, the disjunction of them is also a $G$-clock constraint. $\circled{1}$ is a $G$-clock constraint as assumed. $\circled{2}$ is a $G$-clock constraint because $g$ and $I_G(l')$ are $G$-clock constraints by definition, and $N^i(l')[r]$ is a $G$-clock constraint since $N^i(l')$ is a $G$-clock constraint as assumed, and the reset update represents a $G$-clock constraint according to Lemma~\ref{lemma:reset}. Then, the conjunction of $g$, $I_G(l')$, and $N^i(l')[r]$ gives a $G$-clock constraint by definition. Finally, the big disjuction in $\circled{2}$ is over a finite number of $G$-clock constraints as the number of edges is finite.
$\circled{3}$ is a $G$-clock constraint because
$I_G(l)$ is a $G$-clock constraint by definition, and $N^i(l')$ is a $G$-clock constraint as assumed.
So, based on Lemma~\ref{lemma:Delta-NBP}, $\circled{3}$ represents a $G$-clock constraint.

\textbf{Conclusion:} By the principle of induction, the claim of Lemma~\ref{lemma:NB is cc} holds for any iteration $i$ and location $l\in L$.
\hfill$\blacksquare$
\end{proof}

\begin{lemma}[$\Delta$-Time Invariance for $\BSP$]
\label{lemma:Delta-BSP}
Given $G$-clock constraints $\varphi_1$, $\varphi_2$, and $\varphi_3$,  $\exists\Delta\; \varphi_1^{\uparrow\Delta}\, \wedge \forall\delta \leq \Delta\; \big(\varphi_2^{\uparrow\delta} \wedge 	\forall\delta' \leq \delta\;
\varphi_3^{\uparrow\delta'}\big)$ also represents a $G$-clock constraint.
\end{lemma}


\begin{proof}
Let us indicate $\exists\Delta\; \varphi_1^{\uparrow\Delta}\, \wedge \big(\forall\delta \leq \Delta\; \varphi_2^{\uparrow\delta} \wedge 	\forall\delta' \leq \delta\;
\varphi_3^{\uparrow\delta'}\big)$ by $\Phi$.
Take a clock valuation $u_1$ represented by a clock region of $G$, say $r_G\in R_G$, such that $u_1\models \Phi$.
According to Lemma~\ref{lemma:G-cc}, it suffices to prove that for any region $r_G$ and any two clock valuations $u_1, u_2\in r_G$: $u_1 \models \Phi$ iff $u_2\models \Phi$. Because of symmetry considerations it suffices to prove that $u_1 \models \Phi$ implies $u_2 \models \Phi$.

Consider an arbitrary region $r_G \in R_G$ and arbitrary clock valuations $u_1, u_2 \in r$.
Let us assume $u_1\models \Phi$. Then there exists some $\Delta_1$ such that $u_1\models \varphi_1^{\uparrow\Delta_1}$, and   $u_1\models\forall\delta\leq \Delta_1\; \varphi_2^{\uparrow\delta}\wedge \forall\delta'\leq \delta\; \varphi_3^{\uparrow\delta'}$.
It is proved that there always exists a $\Delta_2$ for which $u_2\models \varphi_1^{\uparrow\Delta_2}$, and $u_2\models\forall\delta\leq \Delta_2\; \varphi_2^{\uparrow\delta}\wedge \forall\delta'\leq \delta\; \varphi_3^{\uparrow\delta'}$.
Let us say that $u_1^{\uparrow\Delta_1}\in r_{\Delta}$ for some $r_{\Delta}\in R_G$. Then, based on Lemma~\ref{lemma:CV}, there exists a $\Delta_2\in\RealTime$ such that $u_2^{\uparrow\Delta_2}\in r_{\Delta}$.
Since $u_1^{\uparrow{\Delta_1}}\models \varphi_1$ and $u_1^{\uparrow{\Delta_1}},u_2^{\uparrow{\Delta_2}}\in r_{\Delta}$, by Lemma~\ref{lemma:G-cc}, we have $u_2^{\uparrow{\Delta_2}}
\models \varphi_1$.

What remains to prove is that for all $\delta\leq\Delta_2$: $u_2\models \varphi_2^{\uparrow\delta}$ and for all $\delta'\leq \delta$: $u_2\models\varphi_3^{\uparrow\delta'}$.
Take $\delta_2\leq \Delta_2$, and assume $u_2^{\uparrow{\delta_2}}\in r_{\delta}$, where $r_{\delta}$ can be $r_G$, $r_{\Delta}$, or any region in between. Then, for any $\delta'\leq \delta_2$, $u_2^{\uparrow\delta'}$ moves to either $r_G$, $r_{\delta}$, or any region in between. Let us take $\delta'_2\leq \delta_2$, and assume that $u_2^{\uparrow\delta'_2}$ moves to $r_{\delta'}$.
Based on Lemma~\ref{lemma:CV}, there exists a $\delta_1$ and a $\delta'_1$ such that $u_1^{\uparrow{\delta_1}}\in r_{\delta}$ and $u_1^{\uparrow{\delta'_1}}\in r_{\delta'_1}$. We prove that $\delta_1\leq \Delta_1$ and $\delta'_1\leq \delta_1$ by contradiction. 1) Assume $\delta_1>\Delta_1$. Then, $r_{\delta}$ already passed $r_\Delta$, and this contradicts the fact that $r_{\delta}$ is either $r_G$, $r_{\Delta}$, or any other region in between. 2) Assume $\delta'_1>\delta_1$. Then, $r_{\delta'}$ already passed $r_\delta$, and this contradicts the fact that $r_{\delta'}$ is either $r_G$, $r_{\delta}$, or any other region in between.
\hfill$\blacksquare$
\end{proof}

\begin{lemma}[Bad State Predicate]
\label{lemma:BS is cc}
Given a plant $G$ and $\NBP(G)$, $B^i(l)$ computed by Algorithm~\ref{algo:BS} in each iteration $i$ and for each location $l\in L$ represents a $G$-clock constraint.
\end{lemma}

\begin{proof}
This is proved in a similar way to the proof of Lemma~\ref{lemma:NB is cc}, where Lemma~\ref{lemma:Delta-BSP} is used to show that $\circled{6}$ represents a $G$-clock constraint.
\hfill$\blacksquare$
\end{proof}

\begin{lemma}[Adapted Guards]
\label{lemma:adapted guards}
Given a plant $G$, $e.g^m$ computed by Algorithm~\ref{algo:synthesis} in each iteration $m$ and for each edge $e\in E_S$ represents a $G$-clock constraint.
\end{lemma}

\begin{proof}
We do the proof by induction on the number of iterations $m$.

\textbf{Base case:} $m=0$. Then, $e.g^m=e.g$ for any $e\in E_S$ which is a $G$-clock constraint by definition.

\textbf{Induction step:} Assume that the statement holds for $m$, i.e., $e.g^m$ is a $G$-clock constraint for all $e\in E_S$. It suffices to prove that the statement holds for $m+1$, i.e., $e.g^{m+1}$ represents a $G$-clock constraint for all $e\in E_S$. 
Consider Algorithm~\ref{algo:synthesis}-line~\ref{line:g-adapt}, $e.g^{m+1}=e.g^m\wedge\neg B^{n,m}(l')[r]$. Now, $e.g^m$ is a $G$-clock constraint as assumed. $B^{n,m}(l')$ is a $G$-clock constraint because according to the proof of Lemma~\ref{lemma:BS is cc}, in each iteration, the bad state predicate of each location is a $G$-clock constraint. 
Based on Lemma~\ref{lemma:reset}, $B^{n,m}(l')[r]$ represents a $G$-clock constraint, and so due to Lemma~\ref{lemma:neg}, $\neg B^{n,m}(l')[r]$ represents a $G$-clock constraint. 
Then, the conjunction of $e.g^m$, and $\neg B^{n,m}(l')[r]$ gives a $G$-clock constraint by definition.

\textbf{Conclusion:} By the principle of induction, the claim of Lemma~\ref{lemma:adapted guards} holds for any iteration $m$ and edge $e\in E_S$.
\hfill$\blacksquare$
\end{proof}


\begin{lemma}[Adapted Invariants]
\label{lemma:adapted invs}
Given a plant $G$, $I^n_S(l)$ computed by Algorithm~\ref{algo:synthesis} in each iteration $n$ and for each location $l\in L$ represents a $G$-clock constraint.
\end{lemma}

\begin{proof}
This is proved in a similar way to the proof of Lemma~\ref{lemma:adapted guards}.
\hfill$\blacksquare$
\end{proof}

\begin{remark}
As Algorithm~\ref{algo:synthesis} terminates (See Appendix~\ref{proof:TSCS termination}), in the coming proofs, for a given a plant $G$ and $\TSCS(G)$, it is assumed that the outer loop (Loop-2) terminates in $n=N$ iterations, and for each $0\leq n\leq N$, the inner loop (Loop-1 inside Loop-2) terminates in $m=M_n$ iterations. $S^{N,M_N}$ denotes the result of the final iteration, so that $S=S^{N,M_N}$ is the output of $\TSCS(G)$. 
\end{remark}

\begin{lemma}[Synthesis Intermediate Results]
\label{lemma:S^n,m is TA}
Given a plant $G$, the result of $\TSCS(G)$ at any iteration $n,m$  ($n\leq N, m\leq M_n$) is a TA.
\end{lemma}

\begin{proof}
Initially, $S$ is set to $G$, which is a TA. Then, at each iteration over $n,m$, only some of the guards and invariants may change. According to Lemma~\ref{lemma:adapted guards} and Lemma~\ref{lemma:adapted invs}, the adapted guards and invariants are always clock constraints. So, based on Definition~\ref{def:TA}, the result of $\TSCS(G)$ at any iteration $n,m$ is a TA.
\hfill$\blacksquare$
\end{proof}

\begin{remark}
As Lemma~\ref{lemma:S^n,m is TA} holds, in the coming proofs, the result of Algorithm~\ref{algo:synthesis} at iteration $n,m$ ($n\leq N, m\leq M_n$) is assumed to be the TA $S^{n,m}$, represented by the automaton $(C,L,\Sigma_G,E^m_S,L_m,l_0,I^n_S)$, where $E^m_S$ is the set of edges, and $I^n_S$ gives the invariants of $S^{n,m}$. 
\end{remark}



\section{Proofs of Properties and Theorems}

\subsection{Proof of Property~\ref{prop:NBPtermination}}\label{proof:NBPtermination}
Based on Lemma~\ref{lemma:NB is cc}, in each iteration of the algorithm, say $i$, and for any location $l\in L$: $N^i(l)$ represents a $G$-clock constraint.  At line~\ref{line:NB-update}, $N^i(l)$ is adapted to the $G$-clock constraint $N^{i+1}(l)=N^i(l)\vee(\circled{2}\vee\circled{3})$ for all $l\in L$. 
Both $\circled{2}$ and $\circled{3}$ represent $G$-clock constraint as proved in Lemma~\ref{lemma:NB is cc}. So, $Z(N^{i+1}(l))=Z(N^i(l))\cup Z(\circled{2}\vee\circled{3})$ where $Z(\circled{2}\vee\circled{3})\in \mathcal{P}(R_G)$.
Then, if $Z(\circled{2}\vee\circled{3})\subseteq Z(N^i(l))$: $Z(N^{i+1}(l))=Z(N^i(l))$. So, $N^{i+1}(l)=N^i(l)$, and the algorithm terminates (line~\ref{line:NBterminate}).
Otherwise, at least a region $r_G\in R_G$ is added to $Z(N^i(l))$ so that $Z(N^{i+1}(l))=Z(N^i(l))\cup \{r_G\}$. Since $L$ and $R_G$ are both finite, this can occur only finitely 
many times.

\subsection{Proof of Property~\ref{prop:NBP}}
\label{proof:NBP}
This property is proved in two parts:


1) take an arbitrary nonblocking state $(l,u)$, and assume that from $(l,u)$, a marked state can be reached in $j$ transitions.
Since the algorithm terminates and in each iteration (line~\ref{line:NB-update}), the nonblocking condition for a given location $l$ is never strengthened, it always holds that $N^i(l)\Rightarrow N(l)$. So, to conclude that $u\models N(l)$, we prove that $u\models N^i(l)$ for some $i$ by induction on $j$:

\textbf{Base case:} assume that from $(l,u)$, a marked state is reached in $0$ transitions. In other words, $l\in L_m$, and so $N^0(l)=I_G(l)$ by definition. Then, for $i=0$, $u\models N^i(l)$ since the semantic graph only contains states $(l,u)$ for which the clock valuation satisfies the invariant of the location; $u\models I_G(l)$.

\textbf{Induction step:} assume that from $(l,u)$, a marked state is reached in $j+1$ transitions. Also, assume that $(l,u)$ leads to a state, say $(l',u')$, in one transition (this means that from $(l',u')$, a marked state is reached in $j$ transitions), where the statement holds for $(l',u')$ i.e., $u'\models N^i(l')$ for some $i$ (by induction assumption). We prove that $u\models N^{i+1}(l)$.

If  $(l,u)$ moves to $(l',u')$ by an event transition, say $\sigma\in \Sigma$, where this transition is related to an edge, $(l,\sigma,g,r,l')$. Then, based on Definition~\ref{def:SG}, $u\models g$, and $u[r]\models I_G(l')$. Also, $u[r]\models N^i(l')$ since $u'\models N^i(l')$ as assumed and $u'=u[r]$. So, $u\models N^{i+1}(l)$ since $u\models \circled{2}$.
If  $(l,u)$ moves to $(l',u')$ by a time transition, say $\Delta$. Then, $u+\Delta \models N^i(l)$ because based on Definition~\ref{def:SG}, $l'=l$, $u'=u+\Delta$, and $u'\models N^i(l')$ as assumed. Also, for all $\delta\leq \Delta$: $u+\delta\models I_G(l)$ by definition. So, $u\models N^{i+1}(l)$ since $u\models \circled{3}$.

\textbf{Conclusion:} for any state $(l,u)$ in the semantic graph of $G$ that is nonblocking: $u\models N(l)$.

2) take an arbitrary $(l,u)$ for which $u \models N(l)$. Since the algorithm terminates, and in each iteration the nonblocking condition for a given location $l$ is never strengthened, there is always some $i$ such that $u \models N^i(l)$. We prove by induction on $i$ that from $(l,u)$, a marked state is reached:



\textbf{Base case:} assume $u\models N^0(l)$. Then, $N^0(l)$ cannot be \false, and so from $(l,u)$, a marked state is reached (in $0$ transitions) as $l\in L_m$.

\textbf{Induction step:} assume $u\models N^{i+1}(l)$, and the statement holds for $i$
, i.e., for any $(l',u')$ with $u'\models N^i(l')$: from $(l',u')$, a marked state is reached (induction assumption).



Considering the nonblocking predicate computation (line\ref{line:NB-update}), $u\models N^{i+1}(l)$ either because already $u\models N^i(l)$, or because $u\models \circled{2}$ or $u\models \circled {3}$.

In case $u\models N^i(l)$. Then, from $(l,u)$, a marked state is reached based on the induction assumption. 

In case $u\models \circled{2}$, then there exists at least one edge $(l,\sigma,g,l',r)$ such that $u\models g \wedge I_G(l')[r]\wedge N^i(l')[r]$.
Since $u\models N^i(l')[r]$ and $u'=u[r]$, $u'\models N^i(l')$. So, based on the induction assumption, from $(l',u')$, a marked state is reached.
Also, since $u\models g \wedge I_G(l')[r]$, due to Definition~\ref{def:SG}, there is an event transition leading from $(l,u)$ to $(l',u')$. So, from $(l,u)$, a marked state is reached. 

In case $u\models \circled{3}$, then there exists $\Delta$ such that $u+\Delta \models N^i(l)$, and for all $\delta\leq \Delta$: $u+\delta \models I_G(l)[r]$. 
Since $u+\Delta \models N^i(l)$  with $u'=u+\Delta$, based on induction assumption, from $(l',u')$, a marked state is reached.
Also, due to Definition~\ref{def:SG}, there is a time transition from $(l,u)$ to $(l',u')$ as for all $\delta\leq \Delta$: $u+\delta \models I_G(l)[r]$. So, from $(l,u)$, a marked state is reached.

\textbf{Conclusion:} from any state $(l,u)$ in the semantic graph of $G$ with $u\models N(l)$, a marked state is reached.

\subsection{Proof of Property~\ref{prop:BSPtermination}}
\label{proof:BSPtermination}

This property is proved in a similar way to the proof of Property~\ref{prop:NBPtermination}, where Lemma~\ref{lemma:Delta-BSP} is used to show that $\circled{6}$ represents a $G$-clock constraint.

\subsection{Proof of Property~\ref{prop:BSP}}
\label{proof:BSP}

This property is proved in two parts:

1) take an arbitrary bad state $(l,u)$, and assume that from $(l,u)$ a blocking state can be reached in $j$ (uncontrollable) transitions. Since the algorithm terminates, and in each iteration (line~\ref{line:BS-update}), the bad state condition for a given location $l$ is never strengthened, it always holds that $B^i(l)\Rightarrow B(l)$. So, to conclude that $u\models B(l)$, we prove that $u\models B^i(l)$ for some $i$ by induction on $j$:



\textbf{Base case:} from $(l,u)$, a blocking state is reached in $0$ transitions. 
Then, due to Property~\ref{prop:NBP} $u\not\models N(l)$, and so for $i=0$, $u\models B^i(l)$ by definition.

\textbf{Induction step:} from $(l,u)$, a blocking state can be reached in $j+1$ (uncontrollable) transitions. 
Assume that in one (uncontrollable) transition, $(l,u)$ moves to a state, say $(l',u')$, where the statement holds for $(l',u')$ i.e., $u'\models B^i(l')$ for some $i$ (by the induction assumption). We prove that $u\models B^{i+1}(l)$.



If $(l,u)$ moves to $(l',u')$ by an uncontrollable event transition that is related to an edge $(l,\sigma,g,r,l')$. Then, based on Definition~\ref{def:SG}, $u\models g$, and $u[r]\models I_G(l')$. Also, $u[r]\models B^i(l')$ since $u'\models B^i(l')$ as assumed and $u'=u[r]$. So, $u\models B^{i+1}(l)$ since $u\models \circled{5}$.

If $(l,u)$ moves to $(l',u')$ by a time transition, say $\Delta$, that is not preemptable. Then, $u\models B^{i+1}(l)$ since $u\models \circled{6}$ for the following reasons: 1) $u+\Delta \models B^i(l)$ because based on Definition~\ref{def:SG}, $l'=l$, $u'=u+\Delta$, and $u'\models B^i(l')$ as assumed, 2) for all $\delta\leq \Delta$: $u+\delta\models I_G(l)$ by definition, and 3) since $\Delta$ is not a preemptable time transition, there is no forcible event enabled at $(l,u)$ so that the condition on forcible events always holds (is $\true$).

\textbf{Conclusion:} for any bad state $(l,u)$ in (the semantic graph of) $G$: $u\models B(l)$.




Assume that $u\models B^i(l)$. We prove by induction on $i$ that from $(l,u)$, a blocking state is reached within $i$ uncontrollable transitions:

\textbf{Base case:} $u\models B^0(l)$. Then, $u\not\models N(l)$ by definition. So, based on Property~\ref{prop:NBP}, $(l,u)$ is not a marked state, and any transition enabled at $(l,u)$ does not lead to a nonblocking state. So, from $(l,u)$, a blocking state is reached in $0$ uncontrollable transitions.

\textbf{Induction step:} assume $u\models B^{i+1}(l)$, where the statement holds for $i$,
i.e., for any $(l',u')$ with $u'\models B^i(l')$: from $(l',u')$, a blocking state is by the induction assumption reached within $i$ uncontrollable transitions.


Considering the bad state predicate computation, $u\models B^{i+1}(l)$ because already $u\models B^i(l)$, or because $u\models \circled{5}$ or  $u\models \circled{6}$.


If $u\models B^i(l)$, then, $(l,u)$ is a bad state based on the induction assumption. 

If $u\models \circled{5}$, then there exists at least one edge $(l,\sigma,g,l',r)$, labeled by an uncontrollable event, such that $u\models g \wedge I_G(l')[r] \wedge B^i(l')$.
Since $u\models B^i(l')[r]$ and $u'[r]=u$, $u'\models B^i(l')$. So, based on the induction assumption, $(l',u')$ is a bad state.

Also, since $u\models g \wedge I_G(l')[r]$, due to Definition~\ref{def:SG}, there is an uncontrollable event transition from $(l,u)$ to $(l',u')$. So, $(l,u)$ is a bad state.

If $u\models \circled{6}$, then there exists $\Delta$ such that $u+\Delta\models B^i(l)$, and for all $\delta\leq \Delta$: $u+\delta\models I_G(l)[r]$ (note that there is no forcible event that can preempt time). 
Since $u+\Delta \models B^i(l)$  with $u'=u+\Delta$, based on the induction assumption, $(l',u')$ is a bad state.

Also, since $u+\delta\models I_G(l)[r]$ for all $\delta\leq \Delta$, due to Definition~\ref{def:SG}, there is a time transition from $(l,u)$ to $(l',u')$ that is not preemptable. So, $(l,u)$ is a bad state.

\textbf{Conclusion:} for any state $(l,u)$ in (the semantic graph of) $G$ such that $u\models B(l)$: $(l,u)$ is a bad state.

\subsection{Proof of Property~\ref{prop:TSCS termination}}
\label{proof:TSCS termination}

Inside each iteration over $n$ (loop-2), the iteration over $m$ (loop-1) terminates because the computation of both $N^{n,m}$ and $B^{n,m}$ terminate due to property~\ref{prop:NBPtermination} and property~\ref{prop:BSPtermination}, respectively. Also, whenever all the guards stay the same (line~\ref{line:loop1-e}).
Due to Lemma~\ref{lemma:adapted guards}, in each iteration $m$ and for any edge $e\in E_S$, $e.g^m$ represents a clock constraint which is adapted to the clock constraint $e.g^{m+1}=e.g^m\wedge \neg B^{n,m}(l')[r]$ at line~\ref{line:g-adapt}. 
Based on the properties stated for $Z$, $Z(e.g^{m+1})=Z(e.g^m)\cap Z(\neg B^{n,m}(l')[r])$, and so $Z(e.g^{m+1})\subseteq Z(e.g^m)$ for all $e\in E_S$.
In case that $Z(e.g^{m+1})=Z(e.g^m)$ for all $e\in E_S$, the iteration over $m$ terminates because $e.g^{m+1}=e.g^m$ for any $e\in E_S$. Otherwise, in each iteration, at least one region $r_G\in R_G$ is excluded from $Z(e.g^m)$ for some $e\in E_S$, i.e., $Z(e.g^{m+1})=Z(e.g^m)\setminus \{r\}$ such that $r\notin Z(e.g^m)$, and so loop-1 can iterate only finitely often as $E_S$ and $R_G$ are both finite.
The iteration over $n$ (loop-2) terminates whenever all location invariants stay the same (line~\ref{line:loop2-e}). Based on Lemma~\ref{lemma:adapted invs}, in each iteration of the algorithm $n$, and for any location $l\in L$: $I^n_S(l)$ represents a clock constraint which is adapted to $I^{n+1}_S=I^n_S(l)\wedge\neg B^{n,m}(l)$ at line~\ref{line:invariant}.
Then, for the same reason stated for termination of loop-1, loop-2 also terminates.

\subsection{Proof of Property~\ref{prop:S is TA}}
\label{proof:S is a TA}
The proof follows immediately from Lemma~\ref{lemma:S^n,m is TA}.

\subsection{Proof of Property~\ref{property:subG}}
\label{proof:subG}
According to Lemma~\ref{lemma:S^n,m is TA}, for any $n,m$  ($n\leq N, m\leq M_n$): $S^{n,m}$ is a TA. Also, according to Property~\ref{prop:S is TA}, $S$ is a TA.
To conclude that $S\subseteq G$, we prove that for all $n\leq N$, for all $m\leq M_n$: $S^{n,m}\subseteq G$
using nested induction on $n$ and $m$. Then, in particular $S^{N,M_N} = \TSCS(G) \subseteq G$, which is to be proven. Induction on $n$:

\textbf{Base case:} $n=0$, and we prove that for all $m\leq M_0$: $S^{0,m}\subseteq G$ by induction on $m$:
\begin{itemize}
\item \textbf{Base case:} $S^{0,0}=G$, and so $S^{0,0}\subseteq G$.

\item \textbf{Induction step:} assume $S^{0,m}\subseteq G$. Then, $S^{0,m+1}$ differs from $S^{0,m}$ only in terms of guards.
So, considering Definition~\ref{def:subA} and the construction of $S$ in Algorithm~\ref{algo:synthesis}, it only suffices to prove that for all $(l_s,\sigma,g^{m+1}_S,r,l_t)\in E^{m+1}_G$: $(l_s,\sigma,g_G,r,l_t)\in E_G$ for some $g_G$ such that $g^{m+1}_S\Rightarrow g_G$.

Take arbitrary edge $(l_s,\sigma,g^{m+1}_S,r,l_t)\in E^{m+1}_G$. Then, considering line~\ref{line:g-adapt}, $(l_s,\sigma,g^{m}_S,r,l_t)\in E^{m}_S$ such that either $g^{m}_S=g^{m+1}_S$, or it is strengthened, and so $g^{m+1}_S\Rightarrow g^{m}_S$. Also, since $S^{0,m}\subseteq G$, then for $(l_s,\sigma,g^{m}_S,r,l_t)\in E^{m}_S$: $(l_s,\sigma,g_G,r,l_t)\in E_G$ for some $g_G$ such that $g^{m}_S\Rightarrow g_G$. Thereto, for all $(l_s,\sigma,g^{m+1}_S,r,l_t)\in E^{m+1}_G$: $(l_s,\sigma,g_G,r,l_t)\in E_G$ for some $g_G$ such that $g^{m+1}_S\Rightarrow g_G$.

\item \textbf{Conclusion:} for all $m \leq M_0$: $S^{0,m}\subseteq G$.
\end{itemize}

\textbf{Induction step:} assume $S^{n,m}\subseteq G$ for all $m \leq M_n$. We prove that $S^{n+1,m}\subseteq G$ for all $m \leq M_{n+1}$ using induction on $m$:

\begin{itemize}
\item \textbf{Base case:} $S^{n,0}\subseteq G$ by assumption. $S^{n+1,0}$ differs from $S^{n,0}$ only in terms of invariants.
So, considering Definition~\ref{def:subA}, it suffices to prove that for all $l\in L$: $I^{n+1}_S(l)\Rightarrow I_G(l)$.
Take arbitrary $l\in L$. Then, $I^{n}_S(l)\Rightarrow I_G(l)$ since $S^{n,0}\subseteq G$.
Considering line~\ref{line:invariant}, at iteration $n+1$, either the invariant stays the same, or it is strengthened such that $I^{n+1}_S(l)\Rightarrow I^{n}_S(l)$. So, $I^{n+1}_S(l)\Rightarrow I_G(l)$ as $I^{n}_S(l)\Rightarrow I_G(l)$.

\item \textbf{Induction step:} assume $S^{n+1,m}\subseteq G$ for all $m\leq M_{n+1}$. Then, $S^{n+1,m+1}$ differs from $S^{n+1,m}$ only in terms of guards. Then, $S^{n+1,m+1}\subseteq G$ for the same reason stated in the previous induction step on $m$.

\item \textbf{Conclusion:} for all $m\leq M_{n}$: $S^{n,m}\subseteq G$.
\end{itemize}

\textbf{Conclusion:} for all $n\leq N$, for all $m \leq M_{n}$: $S^{n,m}\subseteq G$.

\subsection{Proof of Property~\ref{property:base4proofs}}
\label{proof:base4proofs}
Take arbitrary  state $(l,u)$ that is reachable in (the semantic graph of) $S$. We prove that $u\not\models B(l)$ by using induction on the length of the path from $(l_0,u_0)$ to $(l,u)$. 

\textbf{Base case:} $(l,u)=(l_0,u_0)$. Then, we already have assumed that $u_0\not\models B(l_0)$.

\textbf{Induction step:} Assume that $(l,u)$ is reached from a (reachable) state, say $(l',u')$ (by an event or time transition), where the statement holds for $(l',u')$ (by induction assumption), i.e., $u'\not\models B(l')$. We prove that the statement holds for $(l,u)$, i.e., $u\not\models B(l)$ for different cases of transitions from $(l',u')$ to $(l,u)$.

$(l,u)$ is reached from $(l',u')$ by $\sigma\in\Sigma_c$, and assume that this transition is related to an edge $(l',\sigma,g^{M_N},r,l)\in E^{N,M_N}_S$. According to Definition~\ref{def:SG}, $u'\models g^{M_N}$. Also, based on line~\ref{line:g-adapt}, $g^{M_N}$ has been adapted in the last iteration such that
$u'\models \neg B(l)[r]$, which is equivalent to $u'[r] \models \neg B(l)$. Again according to Definition~\ref{def:SG}, $u=u'[r]$, and so $u \models \neg B(l)$, which is equivalent to $u \not \models B(l)$.



$(l,u)$ is reached from $(l',u')$ by a time transition, say $\Delta$, where $F_S(l)\neq \varnothing$. According to Definition~\ref{def:SG}, $u'+\Delta\models I^N(l)$.
Based on line~\ref{line:invariant}, $I^N_S(l)$ has been adapted in the last iteration such that $u'+\Delta \not\models B(l')$. Again according to Definition~\ref{def:SG}, $l'=l$, and $u=u'+\Delta$. So, $u\not \models B(l)$


$(l,u)$ is reached from $(l',u')$ by $\sigma\in\Sigmauc$, and assume that this transition is related to an edge $(l',\sigma,g^{M_N},r,l)\in E^{N,M_N}_S$.
Then, $u'\models g^{M_N}\wedge I^N_S(l)[r]$ by Definition~\ref{def:SG}. By contradiction, assume that $u\models B(l)$. Then, $u'\models B(l)[r]$ because $u'=u[r]$. So, $u'\models B(l')$ as $u'\models \circled{5}$ in the bad state predicate computation of $l'$. $u'\models B(l')$ contradicts the induction assumption, and consequently it must be the case that $u \not\models B(l)$ as required. 

$(l,u)$ is reached from $(l',u')$ by a time transition, say $\Delta$, where $F_S(l)=\varnothing$. Then, Also, $u'+\delta\models I^N_S(l)$ for all $\delta\leq\Delta$ by Definition~\ref{def:SG}. By contradiction, assume that $u\models B(l)$. Then $u'+\Delta\models B(l)[r]$ because $u'=u+\Delta$. Since $F_S(l)=\varnothing$, the condition on $\delta'$ in the bad state predicate computation of $l'$ always gives true.
As a result, $u'\models B(l')$ as $u'\models \circled{6}$ in the bad state predicate computation. $u'\models B(l')$ contradicts the induction assumption, and consequently it must be the case that $u \not\models B(l)$ as required.



\textbf{Conclusion:} for any reachable state $(l,u)$ (in the semantic graph) of $S$: $u\not\models B(l)$.

\subsection{Proof of Theorem~\ref{theorem:controllability}}
\label{proof:controllability}

We need to prove that for any $w \in L(S||G)$ and $\sigma \in \Sigmauc \cup \mathbb{R}_{\geq 0}$, whenever $w \sigma \in L(G)$, then $w\sigma \in L(S||G)$, or $\sigma \in \mathbb{R}_{\geq 0}$ and $w \sigma'\in L(S||G)$ for some $\sigma'\in \SigmaFor$. Consider arbitrary $w \in L(S||G)$ and $\sigma \in \Sigmauc \cup \mathbb{R}_{\geq 0}$, and assume that $w \sigma \in L(G)$. Now assume that $\sigma \not\in \mathbb{R}_{\geq 0}$ or $w \sigma' \not\in L(S||G)$ for all $\sigma'\in \SigmaFor$. It suffices to prove that $w \sigma \in L(S||G)$. Since $S \subseteq G$ (based on Property~\ref{property:subG}), it suffices to prove $w \sigma \in L(S)$.


To conclude that $w\sigma \in L(S)$, we prove that for all $n\leq N$, for all $m\leq M_n$: $w\sigma\in L(S^{n,m})$
using nested induction on $n$ and $m$.
Induction on $n$:

\textbf{Base case:} $n=0$. We prove that for all $m\leq M_0$, $w\sigma \in L(S^{0,m})$ using induction on $m$: 

\begin{itemize}
\item \textbf{Base case:} $w\sigma\in L(S^{0,0})$ since $S^{0,0} = G$ and $w \sigma \in L(G)$ as assumed.

\item \textbf{Induction step:} assume $w\sigma\in L(S^{0,m})$.


$S^{0,m+1}$ may differ from $S^{0,m}$ only because the guards of some edges labeled by controllable events have been modified. Thereto, nothing changes in terms of the occurrence of an uncontrollable event or a time transition so that $w\sigma \in L(S^{0,m+1})$.





\item \textbf{Conclusion:} for all $m\leq M_0$: $w\sigma\in L(S^{0,m})$.

\end{itemize}

\textbf{Induction step:} assume that for all $m \leq M_n$, $w\sigma\in L(S^{n,m})$.
We prove that for all $m \leq M_{n+1}$, $w\sigma\in L(S^{n+1,m})$ using induction on $m$:

\begin{itemize}
\item\textbf{Base case:}
we prove that $w \sigma \in L(S^{n+1,0})$.


Since $w\in L(S||G)$, $w\in L(S)$, and so $w\in L(S^{n,m})$ for any $n,m$ as $S$ is the final result of the algorithm. For $w\in L(S^{n,0})$ (it holds by the induction assumption), assume that there exists some $l_s\in L$ and a clock valuation $u_s$ such that $(l_s,u_s)$ is reached from $(l_0,\bf{0})$ by $w$ in (the semantic graph of) $S^{n,0}$. 
To conclude $w\sigma\in L(S^{n+1,0})$, we prove that $\sigma$ occurs at $(l_s,u_s)$ in $S^{n+1,0}$ for different cases of $\sigma$:

$\sigma\in\Sigmauc$. Based on the assumption, $\sigma$ occurs at $(l_s,u_s)$ in $S^{n,0}$. Assume that $\sigma$ transition is related to an edge $e=(l_s,\sigma,g,r,l_t)$ in $S^{n,0}$. Then, according to Definition~\ref{def:SG}:  $u_s\models e.g^{0}$ and $u_s[r]\models I^{n}_S(l_t)$. 
So, $u_s\models e.g^{0}$, and it suffices to prove that $u_s[r]\models I^{n+1}_S(l_t)$.
We continue the proof by contradiction. Assume that $u_s[r]\not\models I^{n+1}_S(l_t)$.
Then, based on line~\ref{line:invariant}, $u_s[r]\models B^{n,0}(l_t)$ because already $u_s[r]\models I^{n}_S(l_t)$ as assumed.
Considering the computation of the bad state predicate of $l_s$, $u_s\models B^{n,0}(l_s)$ because $u_s\models e.g^{0}\wedge I^{n}_S(l_t)[r]\wedge B^{n,0}(l_t)[r]$.


Since $u_s\models B^{n,0}(l_s)$, based on Property~\ref{prop:BSP}, $(l_s,u_s)$ is a bad state, and this contradicts the assumption that $(l_s,u_s)$ is reachable in $S$ because then due to Property~\ref{property:base4proofs}, $(l_s,u_s)$ is not a bad state. 


$\sigma=\Delta$, and there is no $\sigma'\in\SigmaFor$ such that $w\sigma'\in L(S||G)$. Then, for $w\sigma\in L(S^{n,0})$, according to Definition~\ref{def:SG}: $u_s+\delta\models I^{n}_S(l_s)$ for all $\delta\leq \Delta$. Also, the algorithm does not change the invariant as $F_S(l_s)=\varnothing$ so that $I^{n+1}_S(l_s)=I^{n}_S(l_s)$. So, $\sigma$ occurs at $(l_s,u_s)$ in $S^{n+1,0}$.   

\item \textbf{Induction step:} assume $w\sigma\in L(S^{n+1,m})$.
Then, $w\sigma\in L(S^{n+1,m+1})$ for the same reason stated in the previous induction step on $m$.

\item \textbf{Conclusion:} for all $m\leq M_n$, $w\sigma\in L(S^{n,m})$.
\end{itemize}

\textbf{Conclusion:} for all $n\leq N$, for all $m\leq M_n$: $w\sigma\in L(S^{n,m})$.

\subsection{Proof of Theorem~\ref{theorem:NBness}}
\label{proof:NBness}



First of all $L(S)\subseteq L(G)$, and so $L(S||G)=L(S)$.
So, it suffices to prove that $S$ is nonblocking, i.e., any reachable state in (the semantic graph of) $S$ is nonblocking.


Take arbitrary state $(l,u)$ that is reachable in (the semantic graph of) $S$. According to Property~\ref{property:base4proofs}, $u\not\models B^{N,M_N}(l)$. This, based on Property~\ref{prop:BSP}, means that $(l,u)$ is not a bad state in $S^{N,M_N}$ where $S^{N,M_N}=S$ ($S^{N,M_N}$ is the final result of the algorithm). So, $(l,u)$ is not a blocking state in (the semantic graph of) $S$ as $(l,u)$ is not a bad state.

\subsection{Proof of Theorem~\ref{theorem:MPness}}
\label{proof:MPness}

We need to prove that for any other proper supervisor $S'$: $L(S'||G)\subseteq L(S||G)$.
Take arbitrary $w\in L(S'||G)$. We need to prove that $w\in L(S||G)$. Since $L(S)\subseteq L(G)$, it suffices to prove that $w\in L(S)$. We do the proof by induction on the structure of $w$:

\textbf{Base case:} Assume $w=\epsilon$. Then $w\in L(S)$ by definition.


\textbf{Induction step:} Assume $w=v\sigma$ for some $v\in (\Sigma_G\cup\RealTime)^*$ and $\sigma\in \Sigma_G\cup\RealTime$ where the statement holds for $v$, i.e., $v\in L(S)$. It suffices to prove that the statement holds for $v\sigma$, i.e., $v\sigma\in L(S)$.
To conclude that $v\sigma\in L(S)$, we prove that for all $n\leq N$, for all $m \leq M_n$: $v\sigma\in L(S^{n,m})$
using nested induction on $n$ and $m$.
Induction on $n$:





\begin{itemize}[topsep=0pt,partopsep=0pt]
\item \textbf{Base case:} $n=0$. We prove that for all $m \leq M_0$: $v\sigma\in S^{0,m}$ by induction on $m$:

\begin{itemize}[topsep=0pt,partopsep=0pt]
\item \textbf{Base case:}
$v\sigma\in L(S^{0,0})$ because $S^{0,0}=G$, and $v\sigma\in L(G)$ as $v\sigma\in L(S'||G)$ by assumption.

\item \textbf{Induction step:} assume that $v\sigma\in L(S^{0,m})$. It suffices to prove that $v\sigma\in L(S^{0,m+1})$. $S^{0,m+1}$ differs from $S^{0,m}$ only in terms of the guards of (some) controllable edges. So, according to Definition~\ref{def:SG}, $v\sigma\in L(S^{0,m+1})$ for $\sigma\in \Sigmauc\cup\RealTime$ since the guards of uncontrollable edges and the invariants stay the same, and already $v\sigma\in L(S^{0,m})$. 

Let us say that $\sigma\in\Sigma_c$, and assume that for $v\sigma\in L(S^{0,m})$, there exists states $(l_s,u_s)$ (in the semantic graph of $S^{0,m}$) reached by $v$ from the initial state, and $(l_t,u_t)$ reached from $(l_s,u_s)$ by $\sigma$. Then, due to Definition~\ref{def:SG}, $u_s\models e.g^m$, and $u_s[r]\models I^0(l_t)$.
To conclude that $v\sigma\in L(S^{0,m+1})$, it suffices to prove that $u_s\models e.g^{m+1}$ because the invariants stay the same.
Assume that $u_s\not\models e.g^{m+1}$. Then, considering line~\ref{line:g-adapt}, $u_s\models B^{0,m}(l_t)[r]$.
Again by Definition~\ref{def:SG}, $u_s[r]=u_t$. So, $u_t\models B^{0,m}(l_t)$ as $u_s\models B^{0,m}(l_t)[r]$. Then, by Property~\ref{prop:BSP}, $(l_t,u_t)$ is a bad state, and this contradicts the assumption that $S'$ is a proper supervisor as it does not prevent all the bad states to take care nonblockingness and controllability. 

\item \textbf{Conclusion:} for all $m\leq M_0$: $v\sigma\in L(S^{0,m})$.
\end{itemize}

\item \textbf{Induction step:} assume that for all $m\leq M_n$: $v\sigma\in L(S^{n,m})$. We prove that for all $m\leq M_{n+1}$: $v\sigma\in L(S^{n+1,m})$ using induction on $m$.

\begin{itemize}[topsep=0pt,partopsep=0pt]
\item \textbf{Base case:} We need to prove that $v\sigma\in L(S^{n+1,0})$.

For $v\sigma\in L(S^{n,0})$ (it holds by the induction assumption), assume that there exists some $l_s\in L$ and a clock valuation $u_s$ such that $(l_s,u_s)$ is reached from $(l_0,\bf{0})$ by $v$ in (the semantic graph of) $S^{n,0}$. 
To conclude $v\sigma\in L(S^{n+1,0})$, we prove that $\sigma$ occurs at $(l_s,u_s)$ in $S^{n+1,0}$ for different cases of $\sigma$:

$\sigma$ is an event transition, related to an edge $(l_s,\sigma,g,r,l_t)$. Then, according to Definition~\ref{def:SG}:  $u_s\models e.g^{0}$ and $u_s[r]\models I^{n}_S(l_t)$. 
So, $u_s\models e.g^{0}$, and it suffices to prove that $u_s[r]\models I^{n+1}_S(l_t)$.
We continue the proof by contradiction. Assume that $u_s[r]\not\models I^{n+1}_S(l_t)$.
Then, based on line~\ref{line:invariant}, $u_s[r]\models B^{n,0}(l_t)$ because already $u_s[r]\models I^{n}_S(l_t)$ as assumed.
So, $u_t\models B^{n,0}(l_t)$ as $u_s[r]=u_t$ (again by Definition~\ref{def:SG}), and based on Property~\ref{prop:BSP}, $(l_t,u_t)$ is a bad state, and this contradicts the assumption that $S'$ is a proper supervisor.

$\sigma$ is a time transition, say $\Delta$.
Then, for $v\sigma\in L(S^{n,0})$, according to Definition~\ref{def:SG}: $u_s+\delta\models I^{n}_S(l_s)$ for all $\delta\leq \Delta$. Also, $l_t=l_s$, and $u_t=u_s+\Delta$.
It suffices to prove that $u_s+\delta\models I^{n+1}_S(l_s)$ for all $\delta\leq \Delta$.
By contradiction, assume that for some $\delta\leq \Delta$, $u_s+\delta\not\models I^{n+1}_S(l_s)$.
Then, based on line~\ref{line:invariant}, $u_s+\delta\models B^{n,0}(l_s)$ because $u_s+\delta\models I^{n+1}_S(l_s)$. 
In this case, based on Property~\ref{prop:BSP}, $(l_t,u_s+\delta)$ is a bad state. This contradicts the assumption that $S'$ is a proper supervisor because, as a proper supervisor, it should prevent $(l_t,u_s+\delta)$. However, $(l_t,u_s+\delta)$ can be reached through the $\Delta$ transition that occurs in $S'$.

\item \textbf{Induction step:} assume that $v\sigma\in L(S^{n+1,m})$. Then, $v\sigma\in L(S^{n+1,m+1})$ for the same reason states in the previous induction step on $m$.

\item \textbf{Conclusion:} for all $m\leq M_n$: $v\sigma\in L(S^{n,m})$.
\end{itemize}

\item \textbf{Conclusion:} for all $n\leq N$, for all $m\leq M_n$: $v\sigma\in L(S^{n,m})$.

\end{itemize}

\textbf{Conclusion:} by the principle of induction, $w\in L(S||G)$ for all $w\in L(S'||G)$.

\subsection{Proof of Theorem~\ref{theorem:safety}}
\label{proof:safety}
This proof is inspired from the proof of safety in~\cite{TAC-NSC}.
Since $S$ and $G$ has the same event set $\Sigma_G$ and $\Sigma_R\subseteq\Sigma_G$,  $\Sigma_G\cap\Sigma_R=\Sigma_R$.
So, it suffices to prove that if we take any $w\in P_{\Sigma_R}(L({\mathit{S||(G||R^{\bot}}})))$: $w\in L(R)$.
Take $w\in P_{\Sigma_R}(L({\mathit{S||(G||R^{\bot}}})))$, then due to the projection properties, there exists $w'\in L({\mathit{S||(G||R^{\bot}}}))$ such that $P_{\Sigma_R}(w')=w$.
Also, based on Property~\ref{property:subG}, $L(S) \subseteq L(G||R^{\bot})$, and so $w'\in L(G||R^{\bot})$.
Applying the projection on $\Sigma_R$ gives $P_{\Sigma_R}(w')\in L(R^{\bot})$.
For $w\in P_{\Sigma_R}(L({\mathit{S||(G||R^{\bot}}})))\cap L(R^{\bot})$, $w\in L(R)$ since the blocking state $q_d$ added to $G||R$ to make $G||R^{\bot}$ is removed by $S$ as guaranteed by Theorem~\ref{theorem:NBness}.

\bibliographystyle{IEEEtran}
\bibliography{references}

\begin{thebibliography}{10}
\providecommand{\url}[1]{#1}
\csname url@samestyle\endcsname
\providecommand{\newblock}{\relax}
\providecommand{\bibinfo}[2]{#2}
\providecommand{\BIBentrySTDinterwordspacing}{\spaceskip=0pt\relax}
\providecommand{\BIBentryALTinterwordstretchfactor}{4}
\providecommand{\BIBentryALTinterwordspacing}{\spaceskip=\fontdimen2\font plus
\BIBentryALTinterwordstretchfactor\fontdimen3\font minus
  \fontdimen4\font\relax}
\providecommand{\BIBforeignlanguage}[2]{{%
\expandafter\ifx\csname l@#1\endcsname\relax
\typeout{** WARNING: IEEEtran.bst: No hyphenation pattern has been}%
\typeout{** loaded for the language `#1'. Using the pattern for}%
\typeout{** the default language instead.}%
\else
\language=\csname l@#1\endcsname
\fi
#2}}
\providecommand{\BIBdecl}{\relax}
\BIBdecl

\bibitem{ramadge1989control}
P.~J. Ramadge and W.~M. Wonham, ``The control of discrete event systems,''
  \emph{Proceedings of the IEEE}, vol.~77, no.~1, pp. 81--98, 1989.

\bibitem{wonham2015supervisory}
W.~M. Wonham, ``Supervisory control of discrete-event systems,''
  \emph{Encyclopedia of systems and control}, pp. 1396--1404, 2015.

\bibitem{EFAmodeling}
M.~Skoldstam, K.~Akesson, and M.~Fabian, ``Modeling of discrete event systems
  using finite automata with variables,'' in \emph{2007 46th IEEE Conference on
  Decision and Control}.\hskip 1em plus 0.5em minus 0.4em\relax IEEE, 2007, pp.
  3387--3392.

\bibitem{cassandras2009introduction}
C.~G. Cassandras and S.~Lafortune, \emph{Introduction to discrete event
  systems}.\hskip 1em plus 0.5em minus 0.4em\relax Springer Science \& Business
  Media, 2009.

\bibitem{Lin:14}
F.~Lin, ``Control of networked discrete event systems: Dealing with
  communication delays and losses,'' \emph{SIAM Journal on Control and
  Optimization}, vol.~52, no.~2, pp. 1276--1298, 2014.

\bibitem{Rashidinejad18}
A.~Rashidinejad, M.~Reniers, and L.~Feng, ``Supervisory control of timed
  discrete-event systems subject to communication delays and non-{FIFO}
  observations,'' \emph{IFAC-PapersOnLine}, vol.~51, no.~7, pp. 456 -- 463,
  2018, 14th IFAC Workshop on Discrete Event Systems WODES 2018.

\bibitem{Heemels:10}
W.~M.~H. Heemels, A.~R. Teel, N.~Van~de Wouw, and D.~Nesic, ``Networked control
  systems with communication constraints: Tradeoffs between transmission
  intervals, delays and performance,'' \emph{IEEE Transactions on Automatic
  Control}, vol.~55, no.~8, pp. 1781--1796, 2010.

\bibitem{Wonham:94}
B.~A. Brandin and W.~M. Wonham, ``Supervisory control of timed discrete-event
  systems,'' \emph{IEEE Transactions on Automatic Control}, vol.~39, no.~2, pp.
  329--342, 1994.

\bibitem{alur1994theory}
R.~Alur and D.~L. Dill, ``A theory of timed automata,'' \emph{Theoretical
  computer science}, vol. 126, no.~2, pp. 183--235, 1994.

\bibitem{khoumsi2002supervisory}
A.~Khoumsi, ``Supervisory control of dense real-time discrete-event systems
  with partial observation,'' in \emph{Proceedings of the 6th International
  Workshop on Discrete Event Systems (WODES'02)}.\hskip 1em plus 0.5em minus
  0.4em\relax IEEE, 2002, pp. 105--112.

\bibitem{miremadi2015symbolic}
S.~Miremadi, Z.~Fei, K.~{\AA}kesson, and B.~Lennartson, ``Symbolic supervisory
  control of timed discrete event systems,'' \emph{IEEE Transactions on Control
  Systems Technology}, vol.~23, no.~2, pp. 584--597, 2015.

\bibitem{dubey2009discussion}
A.~Dubey, ``A discussion on supervisory control theory in real-time discrete
  event systems,'' \emph{ISIS}, vol.~9, p. 112, 2009.

\bibitem{wong1991control}
H.~Wong-Toi and G.~Hoffmann, ``The control of dense real-time discrete event
  systems,'' in \emph{Proceedings of the 30th IEEE Conference on Decision and
  Control}, 1991, pp. 1527--1528.

\bibitem{tripakis1999fly}
S.~Tripakis and K.~Altisen, ``On-the-fly controller synthesis for discrete and
  dense-time systems,'' in \emph{International Symposium on Formal
  Methods}.\hskip 1em plus 0.5em minus 0.4em\relax Springer, 1999, pp.
  233--252.

\bibitem{maler1995synthesis}
O.~Maler, A.~Pnueli, and J.~Sifakis, ``On the synthesis of discrete controllers
  for timed systems,'' in \emph{Annual Symposium on Theoretical Aspects of
  Computer Science}.\hskip 1em plus 0.5em minus 0.4em\relax Springer, 1995, pp.
  229--242.

\bibitem{asarin1998controller}
E.~Asarin, O.~Maler, A.~Pnueli, and J.~Sifakis, ``Controller synthesis for
  timed automata,'' \emph{IFAC Proceedings Volumes}, vol.~31, no.~18, pp.
  447--452, 1998.

\bibitem{cassez2005efficient}
F.~Cassez, A.~David, E.~Fleury, K.~G. Larsen, and D.~Lime, ``Efficient
  on-the-fly algorithms for the analysis of timed games,'' in
  \emph{International Conference on Concurrency Theory}.\hskip 1em plus 0.5em
  minus 0.4em\relax Springer, 2005, pp. 66--80.

\bibitem{behrmann2007uppaal}
G.~Behrmann, A.~Cougnard, A.~David, E.~Fleury, K.~G. Larsen, and D.~Lime,
  ``Uppaal-tiga: Time for playing games!'' in \emph{International Conference on
  Computer Aided Verification}.\hskip 1em plus 0.5em minus 0.4em\relax
  Springer, 2007, pp. 121--125.

\bibitem{ehlers2017supervisory}
R.~Ehlers, S.~Lafortune, S.~Tripakis, and M.~Y. Vardi, ``Supervisory control
  and reactive synthesis: a comparative introduction,'' \emph{Discrete Event
  Dynamic Systems}, vol.~27, no.~2, pp. 209--260, 2017.

\bibitem{khoumsi2002efficient}
A.~Khoumsi and M.~Nourelfath, ``An efficient method for the supervisory control
  of dense real-time discrete event systems,'' in \emph{Proceedings of the 8th
  International Conference on Real-Time Computing Systems (RTCSA)}, 2002.

\bibitem{tripakis2001analysis}
S.~Tripakis and S.~Yovine, ``Analysis of timed systems using time-abstracting
  bisimulations,'' \emph{Formal Methods in System Design}, vol.~18, no.~1, pp.
  25--68, 2001.

\bibitem{ouedraogo2010setexp}
L.~Ouedraogo, A.~Khoumsi, and M.~Nourelfath, ``Setexp: a method of
  transformation of timed automata into finite state automata,''
  \emph{Real-Time Systems}, vol.~46, no.~2, pp. 189--250, 2010.

\bibitem{ICARCV20}
A.~Rashidinejad, P.~van~der Graaf, and M.~Reniers, ``Nonblocking supervisory
  control synthesis of timed automata using abstractions and forcible events,''
  in \emph{2020 16th International Conference on Control, Automation, Robotics
  and Vision (ICARCV)}.\hskip 1em plus 0.5em minus 0.4em\relax IEEE, 2020, pp.
  1--8.

\bibitem{CIF}
D.~A. van Beek, W.~Fokkink, D.~Hendriks, A.~Hofkamp, J.~Markovski, J.~Van De
  Mortel-Fronczak, and M.~A. Reniers, ``Cif 3: Model-based engineering of
  supervisory controllers,'' in \emph{International Conference on Tools and
  Algorithms for the Construction and Analysis of Systems}.\hskip 1em plus
  0.5em minus 0.4em\relax Springer, 2014, pp. 575--580.

\bibitem{Supremica}
K.~Akesson, M.~Fabian, H.~Flordal, and R.~Malik, ``Supremica-an integrated
  environment for verification, synthesis and simulation of discrete event
  systems,'' in \emph{2006 8th International Workshop on Discrete Event
  Systems}.\hskip 1em plus 0.5em minus 0.4em\relax IEEE, 2006, pp. 384--385.

\bibitem{Rashidinejad:20}
\BIBentryALTinterwordspacing
A.~Rashidinejad, P.~van~der Graaf, M.~Reniers, and M.~Fabian, ``Non-blocking
  supervisory control of timed automata using forcible events,'' in \emph{15th
  International Workshop on Discrete Event Systems (WODES 2020)}.\hskip 1em
  plus 0.5em minus 0.4em\relax IEEE, 2020, accepted. [Online]. Available:
  \url{https://michelreniers.files.wordpress.com/2020/06/wodes20_0055_fi.pdf}
\BIBentrySTDinterwordspacing

\bibitem{Bengtsson2004}
J.~Bengtsson and W.~Yi, \emph{Timed Automata: Semantics, Algorithms and
  Tools}.\hskip 1em plus 0.5em minus 0.4em\relax Springer Berlin Heidelberg,
  2004, pp. 87--124.

\bibitem{alur1999timed}
R.~Alur, ``Timed automata,'' in \emph{International Conference on Computer
  Aided Verification}.\hskip 1em plus 0.5em minus 0.4em\relax Springer, 1999,
  pp. 8--22.

\bibitem{brandin1994supervisory}
B.~A. Brandin and W.~M. Wonham, ``Supervisory control of timed discrete-event
  systems,'' \emph{IEEE Transactions on Automatic Control}, vol.~39, no.~2, pp.
  329--342, 1994.

\bibitem{ouedraogo2011nonblocking}
L.~Ouedraogo, R.~Kumar, R.~Malik, and K.~Akesson, ``Nonblocking and safe
  control of discrete-event systems modeled as extended finite automata,''
  \emph{IEEE Transactions on Automation Science and Engineering}, vol.~8,
  no.~3, pp. 560--569, 2011.

\bibitem{Flordal:07}
H.~Flordal, R.~Malik, M.~Fabian, and K.~{\AA}kesson, ``Compositional synthesis
  of maximally permissive supervisors using supervision equivalence,''
  \emph{Discrete Event Dynamic Systems}, vol.~17, no.~4, pp. 475--504, 2007.

\bibitem{TAC-NSC}
A.~Rashidinejad, M.~Reniers, and M.~Fabian, ``Networked supervisory control
  synthesis of timed discrete-event systems,'' 2020, manuscript submitted for
  publication.

\end{thebibliography}

\begin{IEEEbiography}[{\includegraphics[width=0.9in,height=1.25in,clip,keepaspectratio]{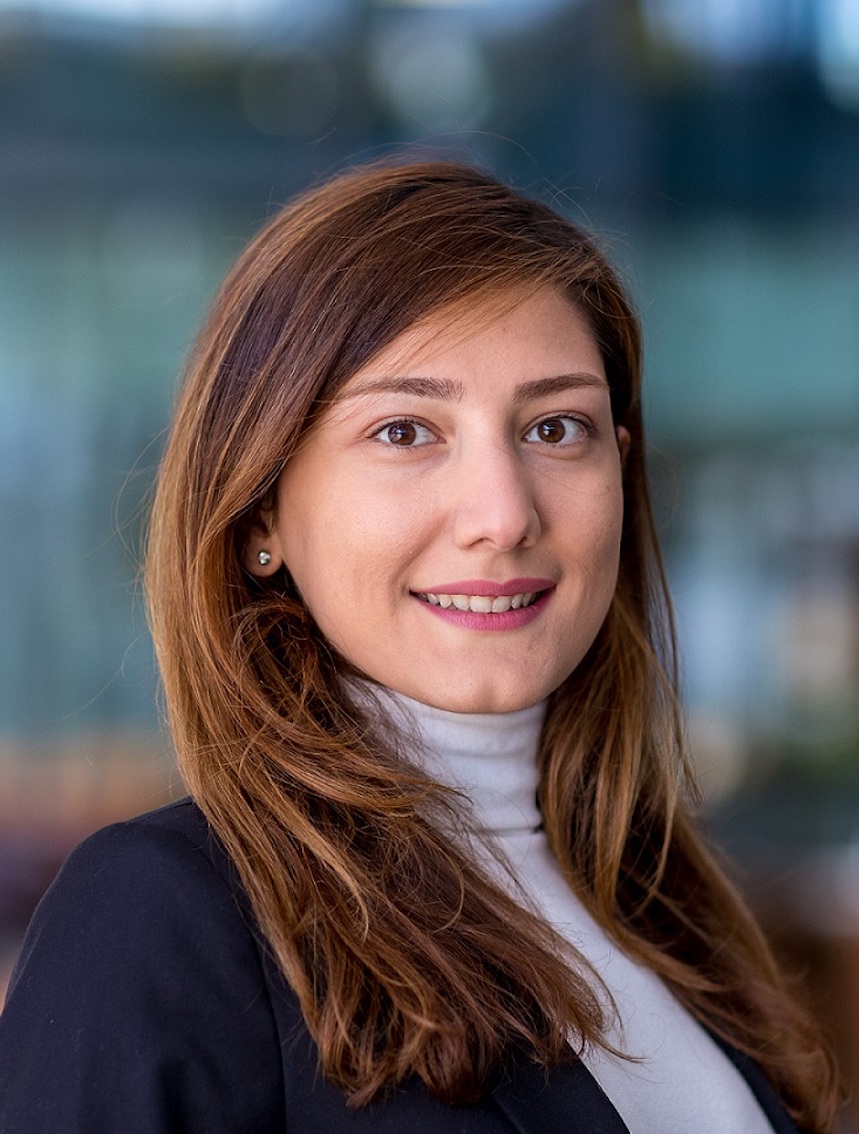}}]{Aida Rashidinejad}
received the M.Sc. degree in electrical-control engineering from Amirkabir University of Technology (Tehran Polytechnic), Tehran, Iran, in 2014. She is currently working towards PhD degree in mechanical engineering-control systems from Eindhoven University of Technology, Eindhoven, The Netherlands. Her current research interests include supervisory control synthesis, networked control, and cyber-physical systems. 
\end{IEEEbiography}

\begin{IEEEbiography}[{\includegraphics[width=1in,height=1.25in,clip,keepaspectratio]{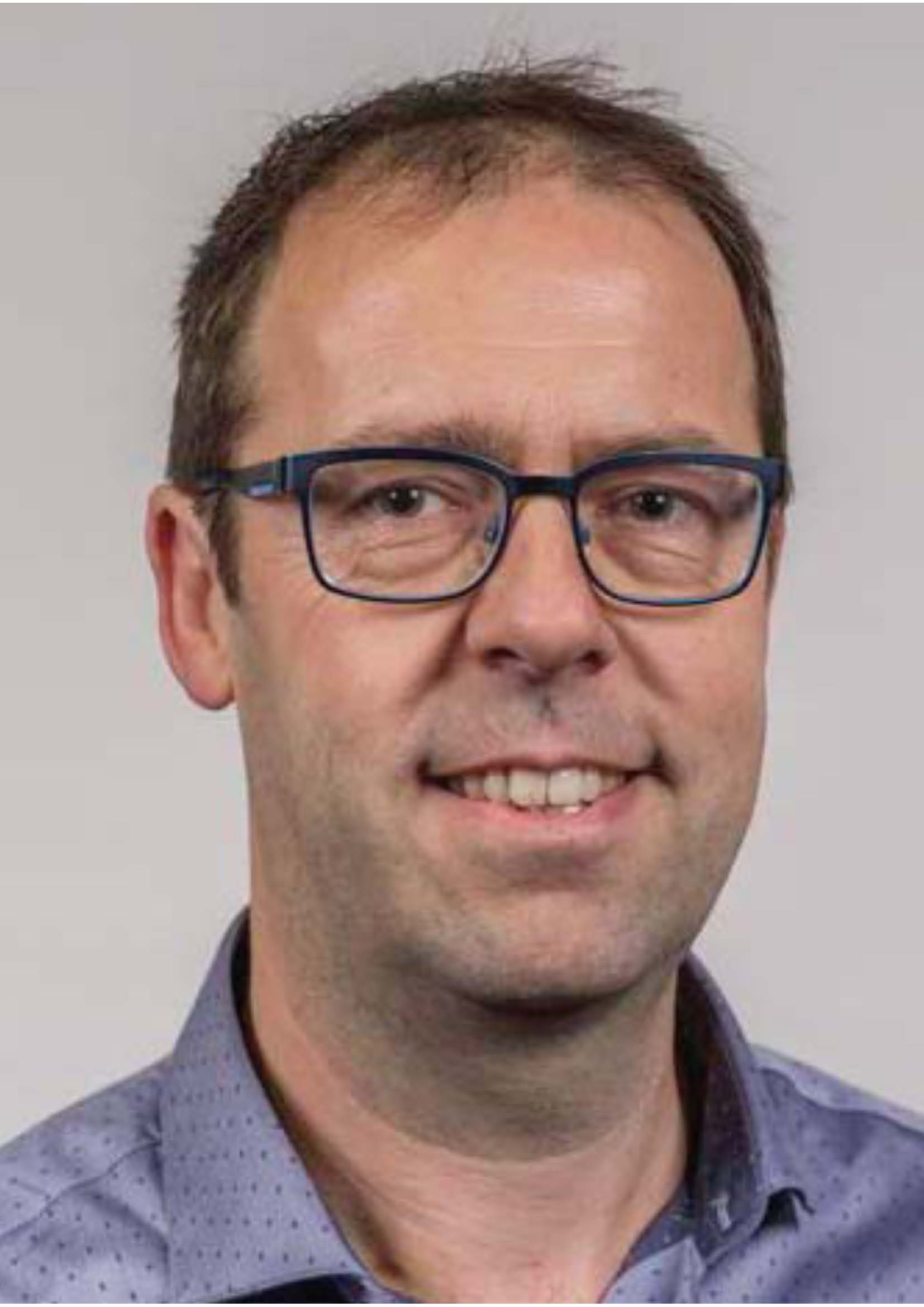}}]{Michel Reniers}
(S’17) is currently an Associate Professor in model-based engineering of
supervisory control at the Department of Mechanical Engineering at TU/e. He has authored
over 100 journal and conference papers. His
research portfolio ranges from model-based systems engineering and model-based validation
and testing to novel approaches for supervisory
control synthesis. Applications of this work are
mostly in the areas of cyber-physical systems.
\end{IEEEbiography}


\begin{IEEEbiography}[{\includegraphics[width=1in,height=1.25in,clip,keepaspectratio]{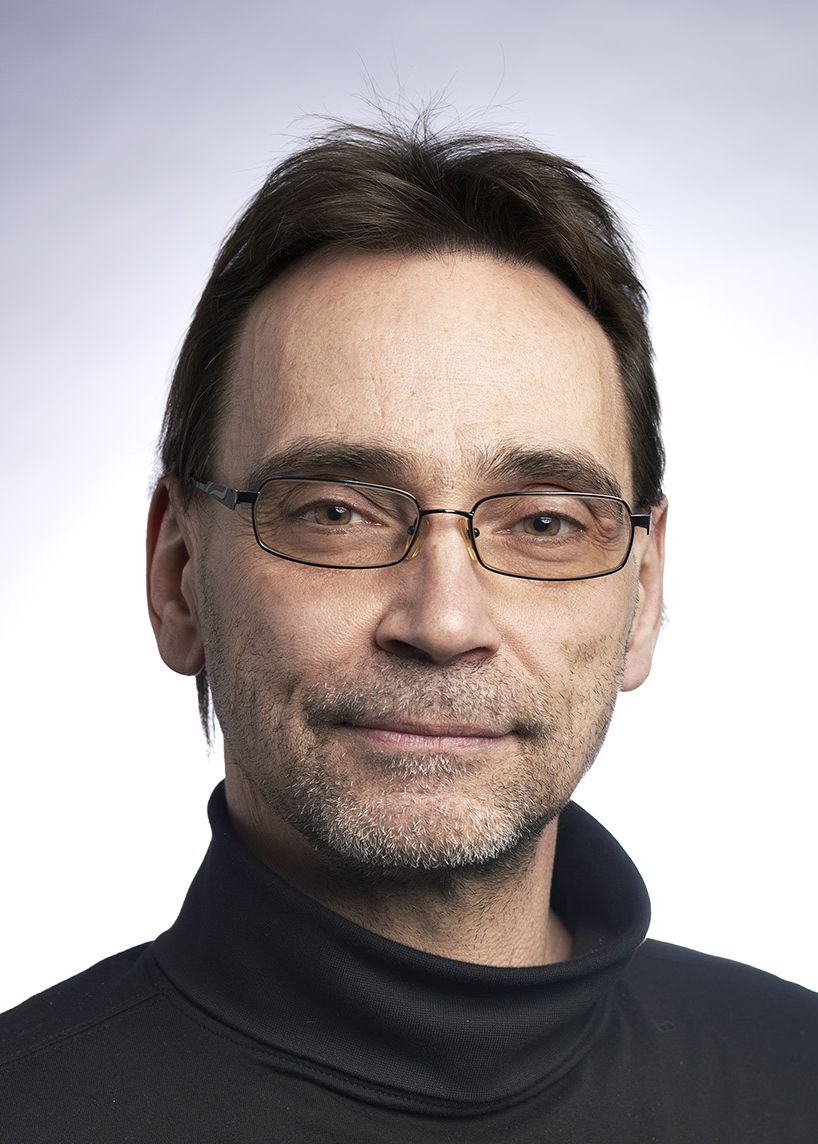}}]{Martin Fabian}
is Professor in Automation and Head of the Automation Research group at the Department of Electrical Engineering, Chalmers University of Technology. His research interests include formal methods for automation systems in a broad sense, merging the fields of Control Engineering and Computer Science. He has authored more than 200 publications, and is co-developer of the formal methods tool Supremica, which implements several state-of-the-art algorithms for supervisory control synthesis.
\end{IEEEbiography}

\end{document}